\documentclass[prd,onecolumn,showpacs,floatfix,amsmath,nofootinbib,amssymb,floatfix]{revtex4}
\usepackage{graphicx,color,dcolumn,booktabs,bm}
\usepackage{longtable,lscape}
\usepackage{txfonts}
\usepackage{overpic,mathrsfs}
\usepackage{amssymb}
\usepackage{indentfirst}
\usepackage{feynmf}   %{feynmp}
\usepackage{slashed}  %for Feynman symbols
\usepackage{cases}
\usepackage{color}
\usepackage{multirow}
\usepackage{epstopdf}
\usepackage[colorlinks,
            citecolor=blue,
            anchorcolor=red,
            menucolor=red,
            linkcolor=red,
            filecolor=red,
            runcolor=red,
            urlcolor=blue,
            frenchlinks=red]{hyperref}

\graphicspath{{Figures/}} %

\allowdisplaybreaks

\begin{document}
%\begin{CJK}{GBK}{}

\title{Searching for possible $\Omega_c-$like molecular states from meson-baryon interaction}

\author{Rui Chen$^{1,2,3}$}
\email{chenr15@lzu.edu.cn}
\author{Atsushi Hosaka$^{3}$}
\email{hosaka@rcnp.osaka-u.ac.jp}
\author{Xiang Liu$^{1,2}$}
\email{xiangliu@lzu.edu.cn}
\affiliation{
$^1$School of Physical Science and Technology, Lanzhou University, Lanzhou 730000, China\\
$^2$Research Center for Hadron and CSR Physics, Lanzhou University
and Institute of Modern Physics of CAS, Lanzhou 730000, China\\
$^3$Research Center for Nuclear Physics (RCNP), Osaka University, Ibaraki, Osaka 567-0047, Japan
}

\begin{abstract}
  Stimulated by the observations of recent observation of $\Omega_c$ as well as former one of $P_c(4380)$ and $P_c(4450)$, we perform a coupled channel analysis of $\Xi_c^*\bar{K}/\Omega_c\eta/\Omega_c^*\eta/\Xi_c\bar{K}^*/\Xi_c'\bar{K}^*/\Omega_c\omega$ systems to search for possible $\Omega_c-$like molecular states by using a one-boson-exchange potential. Our results suggest there exists a loosely bound molecular state, a $\Xi_c^*\bar{K}/\Omega_c\eta/\Omega_c^*\eta/\Xi_c\bar{K}^*/\Xi_c'\bar{K}^*/\Omega_c\omega$ with $I(J^P)=0(3/2^-)$, it is mainly composed of the $\Xi_c^*\bar{K}$ system. Two-body strong decay width is also studied, where we find that $\Xi_c'\bar{K}$ is the dominant decay channel.
\end{abstract}

%\keywords{Molecular State,
%Exotic State, One Pion Exchange, Effective Potential}

\maketitle

\section{Introduction}\label{sec1}

In the past 14 years, a series of novel phenomena relevant to the $XYZ$ states have been reported with the accumulation of experimental data. These observations provide us good chance to identify the exotic configurations (multiquark states, glueball, hybrid) of hadron (see review papers \cite{Chen:2016qju,Liu:2013waa} for more details). In fact, experimental and theoretical studies on
exotic states are a good approach to deepen our understanding of nonperturbative behavior of quantum chromodynamics (QCD).

\begin{figure}[!htbp]
\center
\includegraphics[width=6in]{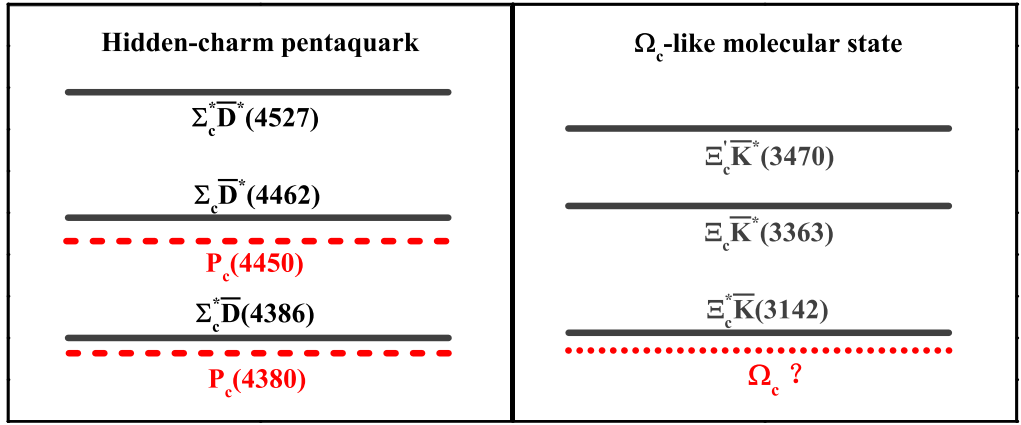}\\
\caption{A comparison of the masses of the heavy pentaquarks and the mass thresholds of a pair anti-meson and baryon systems.}\label{mass}
\end{figure}

In 2015, $P_c(4380)$ and $P_c(4450)$ have been reported by the LHCb Collaboration in the $\Lambda_b^0\to J/\psi pK^-$ decay \cite{Aaij:2015tga}, which stimulated theorists's great enthusiasm of investigating the heavy pentaquarks. As shown in figure \ref{mass}, the masses of $P_c(4380)$ and $P_c(4450)$ are close to the thresholds of a pair of charmed baryon and anti-charmed meson, theorists have tried to describe their inner structures by hadronic molecules \cite{Chen:2015loa,Chen:2015moa,Roca:2015dva,He:2015cea,Burns:2015dwa,Shimizu:2016rrd,Chen:2016heh,Huang:2015uda,Chen:2016otp,Yang:2015bmv,He:2016pfa,Yamaguchi:2016ote,Yamaguchi:2017zmn}. In particular, $P_c(4380)$ was interpreted as a loose molecular pentaquark made up by $\Sigma_c^*\bar{D}$ with several MeV binding energy. Analogously, with a simple replacement of quarks, it is natural to conjecture whether there exist possible $\Omega_c-$like molecules composed of a charmed-strange baryon and an anti-strange meson. In a sense, this study can enhance our understanding of these $P_c$ states.

In fact, the studies of the interaction between one hadron and one strange meson have carried out for several decades. In these works, observations near a pair of hadron thresholds are often explained as possible molecular states as showed in table \ref{study}. Therefore, searching for exotic states from hadron and strange meson interaction is very important, and a further exploration on a charmed-strange baryon and an anti-strange meson interaction is needed.

\renewcommand\tabcolsep{0.8cm}
\renewcommand{\arraystretch}{1.8}
\begin{table}[!htbp]%\tiny
\caption{A summery of the newly observations and their corresponding molecular explanations.\label{study}}
{\begin{tabular}{lll|lll}
\toprule[1pt]
  Observations         &$I(J^P)$      &Explanations              &Observations               &$I(J^P)$      &Explanations    \\\hline
  $X(5568)$            &$1(0^+)$      &$B\bar{K}$ \cite{Xiao:2016mho,Agaev:2016urs,Burns:2016gvy,Albaladejo:2016eps,Chen:2016ypj,Lu:2016kxm,Sun:2016tmz}
  &$D_{s0}(2317)$       &$0(0^+)$      &${D}{K}$ \cite{Barnes:2003dj,Zhang:2009pn,Guo:2006fu,Guo:2011dd,Navarra:2015iea,Xie:2010zza}  \\
  $D_{s1}(2460)$       &$0(1^+)$      &${D}^*{K}$ \cite{Faessler:2007us,Guo:2006rp,Guo:2011dd,Feng:2012zze}
  &$f_0(980)$           &$0(0^+)$      &$K\bar{K}$ \cite{Weinstein:1982gc,Weinstein:1983gd,Weinstein:1990gu} \\
  $f_1(1285)$          &$0(1^+)$      &$K^*\bar{K}$ \cite{Roca:2005nm,Aceti:2015zva,Geng:2015yta,Lu:2016nlp}
  &$\Lambda(1405)$      &$0(1/2^-)$    &$N\bar{K}$ \cite{Dalitz:1959dn,Dalitz:1960du,Ezoe:2016mkp} \\
\bottomrule[1pt]
\end{tabular}}
\end{table}

The LHCb Collaboration announced that five new $\Omega_c$ states were observed in $\Xi_c^+K^-$ channel \cite{Aaij:2017nav}.  {We should emphasize that no resonant-like signals were found in the $\Xi_c^{*+}K^+$ channel, which is a natural case if the observed states are standard three quark states.}
As their masses are very close to the thresholds of a pair of charmed-strange baryon and anti-strange meson, molecular pentaquarks have been proposed in Ref. \cite{Kim:2017jpx,Yang:2017rpg,Huang:2017dwn,Montana:2017kjw,Debastiani:2017ewu}. In this paper, a comprehensive investigation of $\Xi_c^*\bar{K}/\Omega_c\eta/\Omega_c^*\eta/\Xi_c\bar{K}^*/\Xi_c'\bar{K}^*/\Omega_c\omega$ interaction will be adopted, which is also helpful to understand the natures of these new $\Omega_c$ states.

For obtaining the effective potential describing the interaction between the charmed-strange baryon and anti-strange meson, we adopt one-boson-exchange (OBE) model, including $\pi$, $\sigma$, $\rho$, $\omega$ and $\phi$, which is often used to identify and predict exotic states, such as $X/Y/Z/P_c$ states. Both the S-D mixing effect and the coupled channel effect will be also considered here. This study will provide valuable information to search for $\Omega_c-$like molecular states composed of charmed-strange baryon and anti-strange meson. What is more important is that it will also give a test of molecular state picture for $P_c(4380)$ and $P_c(4450)$.

This paper is organized as follows. After introduction, in Sec. \ref{sec2}, the interaction of charmed-strange baryon and anti-strange meson is studied under considered the OBE model and the coupled channel effect. The decay of the possible molecular candidates is further discussed in Sec. \ref{sec3}. The paper ends with a discussion and conclusion in Sec. \ref{sec4}.

\section{The $\Xi_{c}^{(',*)}\bar{K}^{(*)}$ interactions}\label{sec2}

\subsection{Lagrangians}

Let us start with the effective Lagrangian approach in the hadronic level. If considering the heavy quark limit and chiral symmetry \cite{Liu:2011xc}, the effective Lagrangians relevant to the charmed-strange baryon are
\begin{eqnarray}
\mathcal{L}_{\mathcal{B}_{\bar{3}}} &=& l_B\langle\bar{\mathcal{B}}_{\bar{3}}\sigma\mathcal{B}_{\bar{3}}\rangle
          +i\beta_B\langle\bar{\mathcal{B}}_{\bar{3}}v^{\mu}(\mathcal{V}_{\mu}-\rho_{\mu})\mathcal{B}_{\bar{3}}\rangle,\label{lag1}\\
\mathcal{L}_{\mathcal{B}_{6}} &=&  l_S\langle\bar{\mathcal{S}}_{\mu}\sigma\mathcal{S}^{\mu}\rangle
         -\frac{3}{2}g_1\varepsilon^{\mu\nu\lambda\kappa}v_{\kappa}\langle\bar{\mathcal{S}}_{\mu}A_{\nu}\mathcal{S}_{\lambda}\rangle
         +i\beta_{S}\langle\bar{\mathcal{S}}_{\mu}v_{\alpha}\left(\mathcal{V}_{ab}^{\alpha}-\rho_{ab}^{\alpha}\right) \mathcal{S}^{\mu}\rangle
         +\lambda_S\langle\bar{\mathcal{S}}_{\mu}F^{\mu\nu}(\rho)\mathcal{S}_{\nu}\rangle,\\
\mathcal{L}_{\mathcal{B}_{\bar{3}}\mathcal{B}_6} &=& ig_4\langle\bar{\mathcal{S}^{\mu}}A_{\mu}\mathcal{B}_{\bar{3}}\rangle
         +i\lambda_I\varepsilon^{\mu\nu\lambda\kappa}v_{\mu}\langle \bar{\mathcal{S}}_{\nu}F_{\lambda\kappa}\mathcal{B}_{\bar{3}}\rangle+h.c..\label{lag2}
\end{eqnarray}
In the above formula, $A_{\mu}$ and $\mathcal{V}_{\mu}$ are the axial current and vector current, respectively
\begin{eqnarray*}
A_{\mu} = \frac{1}{2}(\xi^{\dag}\partial_{\mu}\xi-\xi\partial_{\mu}\xi^{\dag})=\frac{i}{f_{\pi}}
\partial_{\mu}{P}+\ldots,\quad\quad
\mathcal{V}_{\mu} =
\frac{1}{2}(\xi^{\dag}\partial_{\mu}\xi+\xi\partial_{\mu}\xi^{\dag})
=\frac{i}{2f_{\pi}^2}\left[{P},\partial_{\mu}{P}\right]+\ldots,
\end{eqnarray*}
with $\xi=\text{exp}(i{P}/f_{\pi})$ and the pion decay constant $f_{\pi}=132$ MeV, $\rho_{ba}^{\mu}=ig_V{V}_{ba}^{\mu}/\sqrt{2}$, and $F^{\mu\nu}(\rho)=\partial^{\mu}\rho^{\nu}-\partial^{\nu}\rho^{\mu}
+\left[\rho^{\mu},\rho^{\nu}\right]$. $\mathcal{S}$ is defined as a superfield, which includes $\mathcal{B}_6$ with $J^P=1/2^+$ and $\mathcal{B}^*_6$ with $J^P=3/2^+$ in the $6_F$ flavor representation, i.e.,
$\mathcal{S}_{\mu} =-\sqrt{\frac{1}{3}}(\gamma_{\mu}+v_{\mu})\gamma^5\mathcal{B}_6+\mathcal{B}_{6\mu}^*$. Matrices $\mathcal{B}_{\bar{3}}$, $\mathcal{B}_6^{(*)}$, ${P}$ and ${V}$ are expressed as
\begin{eqnarray*}
\mathcal{B}_{\bar{3}} = \left(\begin{array}{ccc}
        0    &\Lambda_c^+      &\Xi_c^+\\
        -\Lambda_c^+       &0      &\Xi_c^0\\
        -\Xi_c^+      &-\Xi_c^0     &0
\end{array}\right),\quad
\mathcal{B}_6^{(*)} = \left(\begin{array}{ccc}
         \Sigma_c^{{(*)}++}                  &\frac{\Sigma_c^{{(*)}+}}{\sqrt{2}}     &\frac{\Xi_c^{(',*)+}}{\sqrt{2}}\\
         \frac{\Sigma_c^{{(*)}+}}{\sqrt{2}}      &\Sigma_c^{{(*)}0}    &\frac{\Xi_c^{(',*)0}}{\sqrt{2}}\\
         \frac{\Xi_c^{(',*)+}}{\sqrt{2}}    &\frac{\Xi_c^{(',*)0}}{\sqrt{2}}      &\Omega_c^{(*)0}
\end{array}\right),\quad
{P} = {\left(\begin{array}{ccc}
       \frac{\pi^0}{\sqrt{2}} &\pi^+      &K^+\\
       \pi^-       &-\frac{\pi^0}{\sqrt{2}}     &K^0\\
       K^-         &\bar{K}^0      &-\frac{2}{\sqrt{6}}\eta
               \end{array}\right)},\quad
{V} = \left(\begin{array}{ccc}
\frac{\rho^0}{\sqrt{2}}+\frac{\omega}{\sqrt{2}}  &\rho^+      &K^{*+}\\
\rho^- &-\frac{\rho^0}{\sqrt{2}}+\frac{\omega}{\sqrt{2}}      &K^{*0}\\
K^{*-}      &\bar{K}^{*0}    &\phi
\end{array}\right).
\end{eqnarray*}

For the strange meson part, the effective Lagrangians are constructed as \cite{Lin:1999ad,Nagahiro:2008mn}
\begin{eqnarray}
\mathcal{L} &=& \mathcal{L}_{PPV}+\mathcal{L}_{VVP}+\mathcal{L}_{VVV}
             =\frac{ig}{2\sqrt{2}}\langle\partial^{\mu}P\left(PV_{\mu}-V_{\mu}P\right\rangle
               +\frac{g_{VVP}}{\sqrt{2}}\epsilon^{\mu\nu\alpha\beta}\left\langle\partial_{\mu}V_{\nu}\partial_{\alpha}V_{\beta}P\right\rangle
               +\frac{ig}{2\sqrt{2}}\langle\partial^{\mu}V^{\nu}\left(V_{\mu}V_{\nu}-V_{\nu}V_{\mu}\right)\rangle. \label{lag4}
\end{eqnarray}

By expanding Eqs. (\ref{lag1})-(\ref{lag4}), one can further obtain
\begin{eqnarray}
\mathcal{L}_{\mathcal{B}_{\bar{3}}\mathcal{B}_{\bar{3}}\sigma} &=& l_B\langle \bar{\mathcal{B}}_{\bar{3}}\sigma\mathcal{B}_{\bar{3}}\rangle,\\
%%%
\mathcal{L}_{\mathcal{B}_{6}^{(*)}\mathcal{B}_{6}^{(*)}\sigma} &=& -l_S\langle\bar{\mathcal{B}}_6\sigma\mathcal{B}_6\rangle+l_S\langle\bar{\mathcal{B}}_{6\mu}^{*}\sigma\mathcal{B}_6^{*\mu}\rangle
       -\frac{l_S}{\sqrt{3}}\langle\bar{\mathcal{B}}_{6\mu}^{*}\sigma\left(\gamma^{\mu}
       +v^{\mu}\right)\gamma^5\mathcal{B}_6\rangle+h.c.,\\
%%%
\mathcal{L}_{\mathcal{B}_{\bar{3}}\mathcal{B}_{\bar{3}}{V}} &=& \frac{1}{\sqrt{2}}\beta_Bg_V\langle\bar{\mathcal{B}}_{\bar{3}}v\cdot{V}\mathcal{B}_{\bar{3}}\rangle,\\
%%%
\mathcal{L}_{\mathcal{B}_6^{(*)}\mathcal{B}_6^{(*)}{P}} &=&
        i\frac{g_1}{2f_{\pi}}\varepsilon^{\mu\nu\lambda\kappa}v_{\kappa}\langle\bar{\mathcal{B}}_6
        \gamma_{\mu}\gamma_{\lambda}\partial_{\nu}{P}\mathcal{B}_6\rangle
      +i\frac{\sqrt{3}}{2}\frac{g_1}{f_{\pi}}v_{\kappa}\varepsilon^{\mu\nu\lambda\kappa}
      \langle\bar{\mathcal{B}}_{6\mu}^*\partial_{\nu}{P}{\gamma_{\lambda}\gamma^5}
      \mathcal{B}_6\rangle+h.c.
      -i\frac{3g_1}{2f_{\pi}}\varepsilon^{\mu\nu\lambda\kappa}v_{\kappa}\langle
\bar{\mathcal{B}}_{6\mu}^{*}\partial_{\nu} {P}\mathcal{B}_{6\lambda}^*\rangle,\\
%%%
\mathcal{L}_{\mathcal{B}_6^{(*)}\mathcal{B}_6^{(*)} {V}} &=& -\frac{\beta_Sg_V}{\sqrt{2}}\langle\bar{\mathcal{B}}_6v\cdot{V}\mathcal{B}_6\rangle
    -i\frac{\lambda g_V}{3\sqrt{2}}\langle\bar{\mathcal{B}}_6\gamma_{\mu}\gamma_{\nu}
    \left(\partial^{\mu} {V}^{\nu}-\partial^{\nu} {V}^{\mu}\right)
    \mathcal{B}_6\rangle
    -\frac{\beta_Sg_V}{\sqrt{6}}\langle\bar{\mathcal{B}}_{6\mu}^*v\cdot {V}\left(\gamma^{\mu}+v^{\mu}\right)\gamma^5\mathcal{B}_6\rangle\nonumber\\
    &&-i\frac{\lambda_Sg_V}{\sqrt{6}}\langle\bar{\mathcal{B}}_{6\mu}^*
    \left(\partial^{\mu} {V}^{\nu}-\partial^{\nu} {V}^{\mu}\right)
    \left(\gamma_{\nu}+v_{\nu}\right)\gamma^5\mathcal{B}_6\rangle
    +\frac{\beta_Sg_V}{\sqrt{2}}\langle\bar{\mathcal{B}}_{6\mu}^*v\cdot {V}\mathcal{B}_6^{*\mu}\rangle
    +i\frac{\lambda_Sg_V}{\sqrt{2}}\langle\bar{\mathcal{B}}_{6\mu}^*
    \left(\partial^{\mu} {V}^{\nu}-\partial^{\nu} {V}^{\mu}\right)
    \mathcal{B}_{6\nu}^*\rangle+h.c.,\\
\mathcal{L}_{\mathcal{B}_{\bar{3}}\mathcal{B}_6^{(*)}{V}} &=&
       -\frac{\lambda_Ig_V}{\sqrt{6}}\varepsilon^{\mu\nu\lambda\kappa}v_{\mu}\langle \bar{\mathcal{B}}_6\gamma^5\gamma_{\nu}
        \left(\partial_{\lambda} {V}_{\kappa}-\partial_{\kappa} {V}_{\lambda}\right)\mathcal{B}_{\bar{3}}\rangle
          -\frac{\lambda_Ig_V}{\sqrt{2}}\varepsilon^{\mu\nu\lambda\kappa}v_{\mu}\langle \bar{\mathcal{B}}_{6\nu}^*
          \left(\partial_{\lambda}{V}_{\kappa}-\partial_{\kappa}{V}_{\lambda}\right)\mathcal{B}_{\bar{3}}\rangle+h.c.,\\
\mathcal{L}_{\mathcal{B}_{\bar{3}}\mathcal{B}_6^{(*)} {P}} &=& -\sqrt{\frac{1}{3}}\frac{g_4}{f_{\pi}}\langle\bar{\mathcal{B}}_6\gamma^5\left(\gamma^{\mu}+v^{\mu}\right)\partial_{\mu}{P}\mathcal{B}_{\bar{3}}\rangle
           -\frac{g_4}{f_{\pi}}\langle\bar{\mathcal{B}}_{6\mu}^*\partial^{\mu} {P}\mathcal{B}_{\bar{3}}\rangle+h.c.,\\
\mathcal{L}_{\pi KK^*} &=& \frac{ig}{4}\left[\left(\bar{K}^{*\mu}\bm{\tau}\cdot K-\bar{K}\bm{\tau}\cdot K^{*\mu}\right)\partial_{\mu}\bm{\pi}
   +\left(\partial_{\mu}\bar{K}\bm{\tau}\cdot K^{*\mu}-\bar{K}^{*\mu}\bm{\tau}\cdot\partial_{\mu}K\right)\bm{\pi}\right],\\
\mathcal{L}_{\mathbb{V} KK} &=& \frac{ig}{4}\left[\bar{K}\left(\bm{\tau}\cdot\bm{\rho}^{\mu}+{\omega}^{\mu}-\sqrt{2}{\phi}^{\mu}\right)\partial_{\mu}K
     -\partial_{\mu}\bar{K}\left(\bm{\tau}\cdot\bm{\rho}^{\mu}+{\omega}^{\mu}-\sqrt{2}{\phi}^{\mu}\right) K\right],\\
\mathcal{L}_{\mathbb{V} K^*K^*} &=& \frac{ig}{4}
       \left[\left(\bar{K}_{\mu}^*\partial^{\mu}K^{*\nu}-\partial^{\mu}\bar{K}^{*\nu} K_{\mu}^*\right)\left(\bm{\tau}\cdot\bm{\rho}_{\nu}+\omega_{\nu}-\sqrt{2}\phi_{\nu}\right)
       +\left(\partial^{\mu}\bar{K}^{*\nu}K_{\nu}^*-\bar{K}_{\nu}^*\partial^{\mu}K^{*\nu}\right)
       \left(\bm{\tau}\cdot\bm{\rho}_{\mu}+\omega_{\mu}-\sqrt{2}\phi_{\mu}\right)\right.\nonumber\\
       &&\left.+\left(\bar{K}_{\nu}^* K^*_{\mu}-\bar{K}_{\mu}^*K^*_{\nu}\right)
       \left(\bm{\tau}\cdot\partial^{\mu}\bm{\rho}^{\nu}+\partial^{\mu}\omega^{\nu}-\sqrt{2}\partial^{\mu}\phi^{\nu}\right)\right],\\
\mathcal{L}_{\pi K^*K^*} &=& g_{VVP}\varepsilon_{\mu\nu\alpha\beta}
     \partial^{\mu}\bar{K}^{*\nu}\partial^{\alpha}K^{*\beta}\bm{\tau}\cdot\bm{\pi},\\
\mathcal{L}_{\mathbb{V} KK^*} &=& g_{VVP}\varepsilon_{\mu\nu\alpha\beta}
     \left(\partial^{\mu}\bar{K}^{*\nu}K+\bar{K}\partial^{\mu}{K}^{*\nu}\right)\left(\bm{\tau}\cdot\partial^{\alpha}\bm{\rho}^{\beta}+\partial^{\alpha}{\omega}^{\beta}-\sqrt{2}\partial^{\alpha}{\phi}^{\beta}\right).
\end{eqnarray}
Here, $\bm{\tau}$ is Pauli matrix, $\bm{\pi}$ and $\bm{\rho}$ denote the pion and rho meson isospin triplets, respectively. For the $\sigma$ exchange vertex, we use
\begin{eqnarray}
\mathcal{L}_{\sigma K^{(*)}K^{(*)}} &=& g_{\sigma}\bar{K}K\sigma+g_{\sigma}\bar{K}^{*\mu}K^{*}_{\mu}\sigma.
\end{eqnarray}

Coupling constants in above Lagrangians are collected in table \ref{coupling}. The corresponding phase factors between these coupling constants are determined by quark model.
\renewcommand\tabcolsep{0.5cm}
\renewcommand{\arraystretch}{1.5}
\begin{table}[!htbp]
\caption{Coupling constants adopted in our calculation \cite{Liu:2011xc,Lin:1999ad}. $g_{VVP}=3g^2/(32\sqrt{2}\pi^2f_{\pi})$ \cite{Nagahiro:2008mn}.\label{coupling}}
\begin{tabular}{ccccccccccc}
\toprule[1pt]
    $l_B$     &$\beta_Bg_V$     &$l_S$    &$g_1$      &$\beta_Sg_V$      &$\lambda_Sg_V(\text{GeV}^{-1})$      &$g_4$      &$\lambda_Ig_V(\text{GeV}^{-1})$       &$g_{\sigma}$     &$g$\\\hline
     -3.65    &-6.0             &7.3      &1.0        &12.0              &19.2                              &1.06       &-6.8                               &3.65             &12.0\\\bottomrule[1pt]
\end{tabular}
\end{table}

By adopting the Breit approximation, the effective potentials can be related to the scattering amplitudes by
\begin{eqnarray}\label{breit}
\mathcal{V}_{E}^{h_1h_2\to h_3h_4}(\bm{q}) &=&
          -\frac{\mathcal{M}(h_1h_2\to h_3h_4)}
          {\sqrt{\prod_i2M_i\prod_f2M_f}}.
\end{eqnarray}
For this scattering process $h_1h_2\to h_3h_4$, $M_i$ and $M_f$ are the masses of the initial states ($h_1$, $h_2$) and final states ($h_3$, $h_4$), respectively. $\mathcal{M}(h_1h_2\to h_3h_4)$ denotes the scattering amplitude for the $h_1h_2\to h_3h_4$ process by exchanging one boson ($\sigma$, $\pi$, $\rho$, $\omega$ and $\phi$) in $t$-channel. The effective potential in the coordinate space $\mathcal{V}(\bm{r})$ can be obtained through performing Fourier transformation, i.e.,
\begin{eqnarray}
\mathcal{V}_{E}^{h_1h_2\to h_3h_4}(\bm{r}) =
          \int\frac{d^3\bm{q}}{(2\pi)^3}e^{i\bm{q}\cdot\bm{r}}
          \mathcal{V}_{E}^{h_1h_2\to h_3h_4}(\bm{q})\mathcal{F}^2(q^2,m_E^2).\nonumber
\end{eqnarray}
In the above equation, a monopole form factor $\mathcal{F}(q^2,m_E^2)=(\Lambda^2-m_E^2)/(\Lambda^2-q^2)$ is introduced at every interactive vertex to express the off-shell effect of the exchanged boson, where $\Lambda$, $m_E$ and $q$ are the cutoff, mass and four-momentum of the exchanged meson, respectively. %$\Lambda$ is the cutoff with the value around several GeV's.

\subsection{A single channel analysis}\label{single}  %Single $\Xi_c^*\bar{K}$ system

In this subsection, we discuss the single $\Xi_c^*\bar K$ system before performing our full coupled channel study. This is useful to indicate the importance of the coupled channels of $\Xi_c^*\bar K/\Xi_c\bar{K}^*/\Xi_c'\bar{K}^*$.  The OBE effective potential for the S-wave $\Xi_c^*\bar{K}$ system is given as
\begin{eqnarray}
\mathcal{V}_{I}^{\Xi_c^*\bar{K}}(r) &=& -\frac{1}{2}l_Sg_{\sigma}Y(\Lambda,m_{\sigma},r)
                   -\frac{1}{16}\beta_Sg_Vg\left[\mathcal{G}(I)Y(\Lambda,m_{\rho},r)+Y(\Lambda,m_{\omega},r)-2Y(\Lambda,m_{\phi},r)\right], \label{pot1}
\end{eqnarray}
where $\mathcal{G}(I)$ is the isospin factor; 3 for the isoscalar system, and -1 for the isovector system. The corresponding flavor wave functions $|I, I_3\rangle$ for $\Xi_c^*\bar{K}$ system are $|0,0\rangle = (|\Xi_c^{*+}K^{-}\rangle+|\Xi_c^{*0}\bar{K}^{0}\rangle)/\sqrt{2}$, $|1,1\rangle = |\Xi_c^{*+}\bar{K}^{0}\rangle$, $|1,-1\rangle = |\Xi_c^{*0}{K}^{-}\rangle$ and $|1,0\rangle = (|\Xi_c^{*+}K^{-}\rangle-|\Xi_c^{*0}\bar{K}^{0}\rangle)/\sqrt{2}$, respectively. In addition, function $Y(\Lambda,m,{r})$ denotes
\begin{eqnarray}
Y(\Lambda,m,{r}) &=&\frac{1}{4\pi r}(e^{-mr}-e^{-\Lambda r})-\frac{\Lambda^2-m^2}{8\pi \Lambda}e^{-\Lambda r}.\label{yy}
\end{eqnarray}

{{Here the cutoff $\Lambda$ is a parameter of the OBE potential.
As is known well, its reasonable values are around $\Lambda \sim 1$ GeV \cite{Tornqvist:1993ng,Tornqvist:1993vu},
as corresponding to the typical hadronic scale or to the intrinsic size of hadrons.  As we will see in the following sections, in the present paper, we attempt to find bound state solutions by varying the cutoff parameter to obtain binding energy around a few to ten MeV for loosely bound molecular hadrons.  The physical relevance of the results will be discussed in terms of the cutoff parameters.}}

\begin{figure}[!htbp]
\center
\includegraphics[width=6.6in]{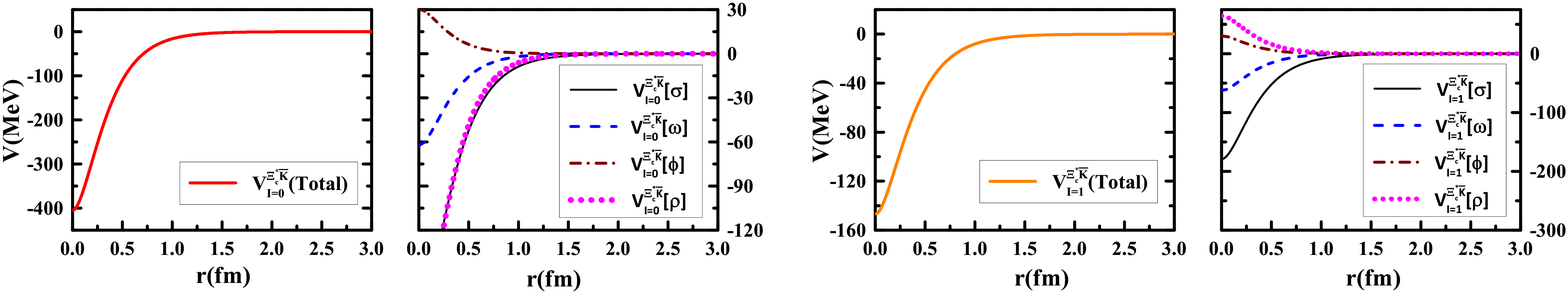}\\
\caption{(color online) $r$ dependence of the deduced effective potentials with cutoff $\Lambda=1.25$ GeV for $\Xi_c^*\bar{K}$ system.}\label{vv}
\end{figure}

In figure \ref{vv}, we present the $r$ dependence of the effective potentials for $\Xi_c^*\bar{K}$ system with $I=0,1$, and its cutoff $\Lambda$ is taken as 1.25 GeV.

According to the effective potential in figure \ref{vv}, we can summarize that:
\begin{itemize}
  \item There does not exist $\pi-$exchange contribution as the spin-parity conservation forbids the vertex $\bar{K}-\bar{K}-\pi$.
  \item The interactions from $\sigma-$exchange is attractive, and it is dominant, which are consistent with the conclusions in Ref. \cite{Chen:2017vai}.
  \item As discussed in Ref. \cite{Chen:2017vai}, the interaction of $\omega-$exchange is different for a hadron-hadron system with light quark-quark or quark-antiquark combinations. For the $\Xi_c^*\bar{K}$ system, its quark configuration is $(csq)-(s\bar{q})$. The $\omega-$exchange from the combination of $q-\bar{q}$ provides attractive interaction.
  \item In comparison with the $\omega-$exchange, $\phi$ meson couples to the strange quark(s) and/or anti-strange quark(s). Here, the $s-\bar{s}$ combination from the $\Xi_c^*\bar{K}$ system implies that the interaction from the $\phi-$exchange is repulsive.
  \item The $\rho$ meson couples to the isospin charge, which explains that in the isoscalar $\Xi_c^*\bar{K}$ system, the interaction from $\rho-$exchange is attractive and almost three times stronger than that in the isovector $\Xi_c^*\bar{K}$ system.
\end{itemize}

By adopting the effective potential in equation (\ref{pot1}), we numerically solve the Schr\"{o}dinger equation, with the cutoff $\Lambda$ being scanned from 1 to 5 GeV. Finally, only the $\Xi_c^*\bar{K}$ state with $[I(J^P)=0(3/2^-)]$ can generate bound solutions when $\Lambda$ is taken larger than 1.60 GeV. As the cutoff $\Lambda$ is increased, it binds deeper and deeper. In particular, when $\Lambda$ is fixed as 2.30 GeV, its binding energy reaches as -21.15 MeV. Then the mass of the $\Xi_c^*\bar{K}$ state with $I(J^P)=0(3/2^-)$ appears close to the observed mass of the newly reported $\Omega_c(3119)$ in \cite{Aaij:2017nav}. However, if we take the cutoff value $\Lambda\sim1$ GeV \cite{Tornqvist:1993ng,Tornqvist:1993vu} as a typical hadron scale and is more reasonable than $\Lambda$ $\sim$ 2 GeV, our results indicate there don't exist any bound solutions for the S-wave $\Xi_c^*\bar{K}$ system, although the total OBE effective potential is attractive. For higher partial waves, due to the repulsive centrifugal potential, it is not likely for bound states exist

To summarize, for the single $\Xi_c^*\bar{K}$ system, the intermediate range and short range forces from the OBE model are not strong enough to generate bound molecular states. Therefor, the newly $\Omega_c(3119)$ cannot be a pure $\Xi_c^*\bar{K}$ molecular state. In Ref. \cite{Debastiani:2017ewu}, the authors proposed that the newly $\Omega_c(3119)$ can be associated with a meson-baryon state strongly coupled to $\Xi_c^*\bar{K}$ and $\Omega_c^*\eta$ with $J^P=3/2^-$ by using the vector meson exchange interaction in the local hidden gauge method. We can find that the vector exchange model provides an attractive interaction, and the coupled channel effect is very important, which is also consistent with our present results: only by considering the single $\Xi_c^*\bar{K}$ system with vector exchange model cannot reproduce the $\Omega_c(3119)$. Thus, in the next section, we will adopt the coupled channel approach to do further study. In our present study, we take into account the long range force from pion exchange, tensor force and S-D wave mixing, which we believe important and was not considered in Ref. \cite{Debastiani:2017ewu}.

\subsection{$\Xi_c^*\bar{K}/\Omega_c\eta/\Omega_c^*\eta/\Xi_c\bar{K}^*/\Xi_c'\bar{K}^*/\Omega_c\omega$ coupled channel system}\label{couple}

In this section, we consider coupled channel studies with the one-pion exchange interaction included, as expected from the nucleon-nucleon interaction. We include the three channels $\Xi_c^*\bar{K}/\Omega_c\eta/\Omega_c^*\eta/\Xi_c\bar{K}^*/\Xi_c'\bar{K}^*/\Omega_c\omega$, and possible quantum numbers are summarized in table \ref{channels}.

\renewcommand\tabcolsep{0.19cm}
\renewcommand{\arraystretch}{1.7}
\begin{table}[!htbp]\center
  \caption{Possible channels involved in our calculation. Here, the first column is the spin-parity quantum numbers corresponding to the channels.}\label{channels}
  {\begin{tabular}{cllllll}
\toprule[1pt]
  \,$J^P$\quad   &\multicolumn{6}{c}{Channels}\\
\hline
$1/2^-$     &$\Xi_c^*\bar{K}:|{}^4\mathbb{D}_{\frac{1}{2}}\rangle$
                    &$\Omega_c\eta:|{}^2\mathbb{S}_{\frac{1}{2}}\rangle$
                    &$\Omega_c^*\eta:|{}^4\mathbb{D}_{\frac{1}{2}}\rangle$
                    &$\Xi_c\bar{K}^*:|{}^2\mathbb{S}_{\frac{1}{2}}/{}^4\mathbb{D}_{\frac{1}{2}}\rangle$
                    &$\Xi_c'\bar{K}^*:|{}^2\mathbb{S}_{\frac{1}{2}}/{}^4\mathbb{D}_{\frac{1}{2}}\rangle$
                    &$\Omega_c\omega:|{}^2\mathbb{S}_{\frac{1}{2}}/{}^4\mathbb{D}_{\frac{1}{2}}\rangle$\\
 $3/2^-$    &$\Xi_c^*\bar{K}:|{}^4\mathbb{S}_{\frac{3}{2}}/{}^4\mathbb{D}_{\frac{3}{2}}\rangle$
                    &$\Omega_c\eta:|{}^2\mathbb{D}_{\frac{3}{2}}\rangle$
                    &$\Omega_c^*\eta:|{}^4\mathbb{S}_{\frac{3}{2}}/{}^4\mathbb{D}_{\frac{3}{2}}\rangle$
                    &$\Xi_c\bar{K}^*:|{}^4\mathbb{S}_{\frac{3}{2}}/{}^2\mathbb{D}_{\frac{3}{2}}/{}^4\mathbb{D}_{\frac{3}{2}}\rangle$
                    &$\Xi_c'\bar{K}^*:|{}^4\mathbb{S}_{\frac{3}{2}}/{}^2\mathbb{D}_{\frac{3}{2}}/{}^4\mathbb{D}_{\frac{3}{2}}\rangle$
                    &$\Omega_c\omega:|{}^4\mathbb{S}_{\frac{3}{2}}/{}^2\mathbb{D}_{\frac{3}{2}}/{}^4\mathbb{D}_{\frac{3}{2}}\rangle$\\
$5/2^-$     &$\Xi_c^*\bar{K}:|{}^4\mathbb{D}_{\frac{5}{2}}\rangle$
                    &$\Omega_c\eta:|{}^2\mathbb{D}_{\frac{5}{2}}\rangle$
                    &$\Omega_c^*\eta:|{}^4\mathbb{D}_{\frac{5}{2}}\rangle$
                    &$\Xi_c\bar{K}^*:|{}^2\mathbb{D}_{\frac{5}{2}}/{}^4\mathbb{D}_{\frac{5}{2}}\rangle$
                    &$\Xi_c'\bar{K}^*:|{}^2\mathbb{D}_{\frac{5}{2}}/{}^4\mathbb{D}_{\frac{5}{2}}\rangle$
                    &$\Omega_c\omega:|{}^2\mathbb{D}_{\frac{5}{2}}/{}^4\mathbb{D}_{\frac{5}{2}}\rangle$\\\hline
$1/2^+$     &$\Xi_c^*\bar{K}:|{}^4\mathbb{P}_{\frac{1}{2}}\rangle$
                    &$\Omega_c\eta:|{}^2\mathbb{P}_{\frac{1}{2}}\rangle$
                    &$\Omega_c^*\eta:|{}^4\mathbb{P}_{\frac{1}{2}}\rangle$
                    &$\Xi_c\bar{K}^*:|{}^2\mathbb{P}_{\frac{1}{2}}/{}^4\mathbb{P}_{\frac{1}{2}}\rangle$
                    &$\Xi_c'\bar{K}^*:|{}^2\mathbb{P}_{\frac{1}{2}}/{}^4\mathbb{P}_{\frac{1}{2}}\rangle$
                    &$\Omega_c\omega:|{}^2\mathbb{P}_{\frac{1}{2}}/{}^4\mathbb{P}_{\frac{1}{2}}\rangle$\\
$3/2^+$     &$\Xi_c^*\bar{K}:|{}^4\mathbb{P}_{\frac{3}{2}}\rangle$
                    &$\Omega_c\eta:|{}^2\mathbb{P}_{\frac{3}{2}}\rangle$
                    &$\Omega_c^*\eta:|{}^4\mathbb{P}_{\frac{3}{2}}\rangle$
                    &$\Xi_c\bar{K}^*:|{}^2\mathbb{P}_{\frac{3}{2}}/{}^4\mathbb{P}_{\frac{3}{2}}\rangle$
                    &$\Xi_c'\bar{K}^*:|{}^2\mathbb{P}_{\frac{3}{2}}/{}^4\mathbb{P}_{\frac{3}{2}}\rangle$
                    &$\Omega_c\omega:|{}^2\mathbb{P}_{\frac{3}{2}}/{}^4\mathbb{P}_{\frac{3}{2}}\rangle$\\
$5/2^+$     &$\Xi_c^*\bar{K}:|{}^4\mathbb{P}_{\frac{5}{2}}\rangle$
                    &$\Omega_c^*\eta:|{}^4\mathbb{P}_{\frac{5}{2}}\rangle$
                    &$\Xi_c\bar{K}^*:|{}^4\mathbb{P}_{\frac{5}{2}}\rangle$
                    &$\Xi_c'\bar{K}^*:|{}^4\mathbb{P}_{\frac{5}{2}}\rangle$
                    &$\Omega_c\omega:|{}^4\mathbb{P}_{\frac{5}{2}}\rangle$\\

\bottomrule[1pt]
\end{tabular}}
\end{table}
The general expressions of spin-orbit wave functions $|{}^{2S+1}L_J\rangle$ for the investigated systems are constructed as
\begin{eqnarray}
\Omega_c\eta:\, \left|{}^{2S+1}L_{J}\right\rangle &=& \sum_{m_S,m_L}C^{J,M}_{\frac{1}{2}m_S,Lm_L}\chi_{\frac{1}{2}m}|Y_{L,m_L}\rangle,\nonumber\\
\Xi_c^*\bar{K}/\Omega_c^*\eta:\, \left|{}^{2S+1}L_{J}\right\rangle &=&
\sum_{m_S,m_L}C^{J,M}_{\frac{3}{2}m_S,Lm_L}
          \Phi_{\frac{3}{2}m_S}|Y_{L,m_L}\rangle,\nonumber\\
\Xi_c^{(')}\bar{K}^*/\Omega_c\omega: \left|{}^{2S+1}L_{J}\right\rangle &=&
\sum_{m,m'}^{m_S,m_L}C^{S,m_S}_{\frac{1}{2}m,1m'}C^{J,M}_{Sm_S,Lm_L}
          \chi_{\frac{1}{2}m}\epsilon^{m'}|Y_{L,m_L}\rangle.\nonumber
\end{eqnarray}
Here, $C^{J,M}_{Sm_S,Lm_L}$, $C^{S,m_S}_{\frac{1}{2}m,1m'}$ and $C^{S,m_S}_{\frac{3}{2}m,1m'}$ are the Clebsch-Gordan coefficients. $\chi_{\frac{1}{2}m}$ and $Y_{L,m_L}$ stand for the spin wave function and the spherical harmonics function, respectively. The polarization vector $\epsilon$ for the vector meson is defined as $\epsilon_{\pm}^{m}=\mp\frac{1}{\sqrt{2}}\left(\epsilon_x^{m}{\pm}i\epsilon_y^{m}\right)$ and $\epsilon_0^{m}=\epsilon_z^{m}$, which satisfy $\epsilon_{\pm1}= \frac{1}{\sqrt{2}}\left(0,\pm1,i,0\right)$ and $\epsilon_{0} =\left(0,0,0,-1\right)$. The polarization tensor $\Phi_{\frac{3}{2}m}$ for baryon $\Xi_c^*$ with spin-3/2 is constructed as $\Phi_{\frac{3}{2}m}=\sum_{m_1,m_2}\langle\frac{1}{2},m_1;1,m_2|\frac{3}{2},m\rangle
\chi_{\frac{1}{2},m_1}\epsilon^{m_2}$.

The total effective potential for the $\Xi_c^*\bar{K}/\Omega_c\eta/\Omega_c^*\eta/\Xi_c\bar{K}^*/\Xi_c'\bar{K}^*/\Omega_c\omega$ system is given in a matrix form,
\begin{eqnarray}
\mathcal{V}_{I,J^P} &=& {\left(\begin{array}{cccccc}
\mathcal{V}_{I,J^P}^{\Xi_c^*\bar{K}\to\Xi_c^*\bar{K}}
              &\mathcal{V}_{I,J^P}^{\Omega_c\eta\to\Xi_c^*\bar{K}}
              &\mathcal{V}_{I,J^P}^{\Omega_c^*\eta\to\Xi_c^*\bar{K}}
              &\mathcal{V}_{I,J^P}^{\Xi_c\bar{K}^*\to\Xi_c^*\bar{K}}
              &\mathcal{V}_{I,J^P}^{\Xi_c'\bar{K}^*\to\Xi_c^*\bar{K}}
              &\mathcal{V}_{I,J^P}^{\Omega_c\omega\to\Xi_c^*\bar{K}}\\
\mathcal{V}_{I,J^P}^{\Xi_c^*\bar{K}\to\Omega_c\eta}
              &\mathcal{V}_{I,J^P}^{\Omega_c\eta\to\Omega_c\eta}
              &\mathcal{V}_{I,J^P}^{\Omega_c^*\eta\to\Omega_c\eta}
              &\mathcal{V}_{I,J^P}^{\Xi_c\bar{K}^*\to\Omega_c\eta}
              &\mathcal{V}_{I,J^P}^{\Xi_c'\bar{K}^*\to\Omega_c\eta}
              &\mathcal{V}_{I,J^P}^{\Omega_c\omega\to\Omega_c\eta}\\
\mathcal{V}_{I,J^P}^{\Xi_c^*\bar{K}\to\Omega_c^*\eta}
              &\mathcal{V}_{I,J^P}^{\Omega_c\eta\to\Omega_c^*\eta}
              &\mathcal{V}_{I,J^P}^{\Omega_c^*\eta\to\Omega_c^*\eta}
              &\mathcal{V}_{I,J^P}^{\Xi_c\bar{K}^*\to\Omega_c^*\eta}
              &\mathcal{V}_{I,J^P}^{\Xi_c'\bar{K}^*\to\Omega_c^*\eta}
              &\mathcal{V}_{I,J^P}^{\Omega_c\omega\to\Omega_c^*\eta}\\
\mathcal{V}_{I,J^P}^{\Xi_c^*\bar{K}\to\Xi_c\bar{K}^*}
              &\mathcal{V}_{I,J^P}^{\Omega_c\eta\to\Xi_c\bar{K}^*}
              &\mathcal{V}_{I,J^P}^{\Omega_c^*\eta\to\Xi_c\bar{K}^*}
              &\mathcal{V}_{I,J^P}^{\Xi_c\bar{K}^*\to\Xi_c\bar{K}^*}
              &\mathcal{V}_{I,J^P}^{\Xi_c'\bar{K}^*\to\Xi_c\bar{K}^*}
              &\mathcal{V}_{I,J^P}^{\Omega_c\omega\to\Xi_c\bar{K}^*}\\
\mathcal{V}_{I,J^P}^{\Xi_c^*\bar{K}\to\Xi_c'\bar{K}^*}
              &\mathcal{V}_{I,J^P}^{\Omega_c\eta\to\Xi_c'\bar{K}^*}
              &\mathcal{V}_{I,J^P}^{\Omega_c^*\eta\to\Xi_c'\bar{K}^*}
              &\mathcal{V}_{I,J^P}^{\Xi_c\bar{K}^*\to\Xi_c'\bar{K}^*}
              &\mathcal{V}_{I,J^P}^{\Xi_c'\bar{K}^*\to\Xi_c'\bar{K}^*}
              &\mathcal{V}_{I,J^P}^{\Omega_c\omega\to\Xi_c'\bar{K}^*}\\
\mathcal{V}_{I,J^P}^{\Xi_c^*\bar{K}\to\Omega_c\omega}
              &\mathcal{V}_{I,J^P}^{\Omega_c\eta\to\Omega_c\omega}
              &\mathcal{V}_{I,J^P}^{\Omega_c^*\eta\to\Omega_c\omega}
              &\mathcal{V}_{I,J^P}^{\Xi_c\bar{K}^*\to\Omega_c\omega}
              &\mathcal{V}_{I,J^P}^{\Xi_c'\bar{K}^*\to\Omega_c\omega}
              &\mathcal{V}_{I,J^P}^{\Omega_c\omega\to\Omega_c\omega}\\\end{array}\right)},\label{hh}
\end{eqnarray}
where subscripts $I$ and $J^P$ are the isospin and spin-parity for the $\Xi_c^*\bar{K}/\Omega_c\eta/\Omega_c^*\eta/\Xi_c\bar{K}^*/\Xi_c'\bar{K}^*/\Omega_c\omega$ system, respectively. The superscript stands for the corresponding scattering process. The concrete expressions for each component shown in equation (\ref{hh}) are presented in section appendix \ref{app01}.

With the preparation of the effective potentials, in the following, we will solve the coupled Schr\"{o}dinger equation to search for bound states, where cutoff $\Lambda$ is scanned from 1 GeV to 5 GeV. Before we show our numerical results, several remarks are in order for the relation between a loosely bound molecular scenario and the cutoff parameter $\Lambda$ introduced in the form factor.
\begin{enumerate}
  \item For a molecular state composed of hadrons $A$ and $B$, its mass is determined as $M=M_A+M_B+E$, here, $E$ is the binding energy. For large $r$ where the interaction is sufficiently suppressed, the asymptotic form of wave function for the $S-$wave $AB$ molecular state has the form $\psi(r)\sim e^{-\sqrt{2\mu E}r}/r$, where $\mu=M_AM_B/(M_A+M_B)$ is the reduced mass. By using the approximate wave function, we find the relation between the size and the binding energy of the molecular state, $R \sim 1/ \sqrt{2\mu E}$. For a loosely bound state, the size of the system should be much larger than the size of all component hadrons, say a few fm or larger. Therefore a good candidate of molecular states should have a binding energy around several MeV or less.

  \item In the coupled channel approach, the binding energy $E$ is measured from the the lowest mass threshold, $M_{th}$, among various channels. Here we introduce another energy $\tilde E$ measured form the mass threshold of the most dominant channel $M_{dom}$, $\tilde E = E + (M_{th} - M_{dom})$. For example, if there is a bound state for the $\Xi_c^*\bar{K}/\Xi_c\bar{K}^*/\Xi_c'\bar{K}^*$ system, and $\Xi_c'\bar{K}^*$ is the dominant channel, $\tilde E = E + (M_{\Xi_c^*\bar{K}} - M_{\Xi_c'\bar{K}^*})$, which may amount to be several hundreds MeV. If $\tilde E \sim 200$ MeV, by using the relation $R\sim 1/\sqrt{2\mu E_{bin}}$, we find that the size $R$ turn out to be about 0.5 fm or even less, which can not be identified with a molecular state.

  \item Because the hadrons are not point-like particles, we introduce a form factor in a monopole manner in every interaction vertex. As a typical hadronic scale, in Ref. \cite{Tornqvist:1993ng,Tornqvist:1993vu}, a reasonable value of cutoff $\Lambda$ should be around 1 GeV.
\end{enumerate}

\renewcommand\tabcolsep{0.42cm}
\renewcommand{\arraystretch}{1.7}
\begin{table}[!hbtp]
\caption{Bound state properties (binding energy $E$ and root-mean-square radius $r_{RMS}$) for all the investigated systems after the coupled channel effects are considered. Probability for the different channels are also given. $E$, $r_{RMS}$, and $\Lambda$ are in units of MeV, fm, and GeV, respectively. To emphasize the dominant channel, we label the probability for the corresponding channel in a bold manner.}\label{num1}
\begin{tabular}{c|ccc|cccccccc}
\toprule[1pt]
 $I(J^{P})$      &$\Lambda$    &$E$       &$r_{RMS}$
      &$\Xi_{c}^*\bar{K}(\%)$   &$\Omega_c\eta(\%)$     &$\Omega_c^*\eta(\%)$      &$\Xi_{c}\bar{K}^*(\%)$     &$\Xi_{c}'\bar{K}^*(\%)$                &$\Omega_c\omega(\%)$   \\\midrule[1pt]
$0(1/2^-)$       &1.182    &-0.53      &0.74                 &15.97      &0.63       &0.75       &$\bm{70.21}$       &11.98     &6.03\\
                         &1.186    &-3.28      &0.71                 &15.63      &0.65       &0.75       &$\bm{70.46}$       &12.06     &6.06\\
                         &1.190    &-6.06      &0.69                 &15.34      &0.66       &0.76       &$\bm{70.67}$       &12.12     &6.10\\

$0(1/2^+)$       &1.709    &-0.26      &0.59                 &1.85      &0.28       &0.19       &$\bm{52.85}$       &18.66     &26.22\\
                         &1.712    &-4.18      &0.50                 &1.16      &0.28       &0.16       &$\bm{55.00}$       &18.94     &26.15\\
                         &1.715    &-8.15      &0.48                 &0.84      &0.28       &0.14       &$\bm{53.03}$       &19.11     &26.59\\\hline

$0(3/2^-)$       &1.270    &-0.58      &5.17                 &$\bm{98.22}$     &$\sim0$    &0.21       &1.50        &$\sim0$   &0.07\\
                         &1.320    &-6.00      &2.17                 &$\bm{92.04}$     &$\sim0$    &0.99       &6.68        &0.03      &0.24\\
                         &1.370    &-20.55     &1.19                 &$\bm{79.27}$     &$\sim0$    &2.48       &27.44       &0.21      &0.50\\
$0(3/2^+)$       &1.945    &-0.16      &1.44              &$\bm{38.93}$      &0.01      &6.57      &$\bm{36.92}$       &14.63       &2.84\\
                         &1.949    &-3.23      &0.95              &$\bm{36.12}$      &0.01      &6.90      &$\bm{38.59}$       &15.39       &2.99\\
                         &1.953    &-6.44      &0.81              &$\bm{34.58}$      &0.01      &7.09      &$\bm{39.39}$       &15.85       &3.07\\\hline

$0(5/2^-)$       &4.480    &-1.98       &0.26             &6.86       &0.09      &4.84      &$\bm{31.52}$       &$\bm{29.67}$       &$\bm{27.01}$\\
                         &4.482    &-6.07       &0.26             &6.88       &0.09      &4.86      &$\bm{31.54}$       &$\bm{29.68}$       &$\bm{26.94}$\\
                         &4.484    &-10.19      &0.26             &6.90       &0.09      &4.88      &$\bm{36.79}$       &$\bm{29.69}$       &$\bm{26.87}$\\
$0(5/2^+)$       &2.096    &-0.63       &0.73             &17.94      &\ldots          &0.48      &$\bm{45.32}$       &26.75       &9.50\\
                         &2.100    &-3.47       &0.58             &17.31      &\ldots          &0.49      &$\bm{45.80}$       &26.82       &9.58\\
                         &2.104    &-6.35       &0.53             &16.99      &\ldots          &0.50      &$\bm{46.10}$       &26.78       &9.63\\
\bottomrule[1pt]
\end{tabular}
\end{table}

According to the above discussions, in table \ref{num1}, results for the binding energies, root-mean-square radii $r_{RMS}$, probabilities of various components of wave functions are shown for all possible spin-parity configurations. There, the binding energies are measured from the lowest threshold, for which we show the results around a few MeV by tuning the cutoff $\Lambda$ to respect the criterion for the molecular state. For each case, we show the results for three different $\Lambda$ values to show possible $\Lambda$ dependence. From the the numerical results presented in table \ref{num1}, we see that:
\begin{itemize}
  \item For $I(J^P)=0({3}/{2}^-)$, when the cutoff $\Lambda$ is taken around 1.30 GeV, there exists one bound state solution with several MeV binding energy and its root-mean-square (RMS) radius larger than 1 fm. Compared to the results for the deuteron \cite{Tornqvist:1993ng,Tornqvist:1993vu}, the $\Xi_c^*\bar{K}/\Xi_c\bar{K}^*/\Xi_c'\bar{K}^*$ state with $I(J^P)=0({3}/{2}^-)$ is a candidate of a molecular state. In this state, $\Xi_c^*\bar{K}|{}^4S_{\frac{3}{2}}\rangle$ channel has dominant contribution, which is more than fifty percent for the $\Xi_c^*\bar{K}/\Omega_c\eta/\Omega_c^*\eta/\Xi_c\bar{K}^*/\Xi_c'\bar{K}^*/\Omega_c\omega$ molecular state. Since this state satisfies the requirements for a loosely bound system, with the size a few fm, and the cutoff around 1 GeV, it is a good candidate of a hadronic molecule with the quantum numbers of $\Omega_c$.

  \item For $I(J^P)=0({1}/{2}^{-})$, it is possible to find a bound state with a binding energy around a few MeV with the cutoff $\Lambda$ around 1 GeV.   However, the resulting bound states have a small size as their corresponding RMS radii are around 0.5 fm. We shall come back to this point shortly.

  \item For $I(J^P)=$ $0({3}/{2}^+)$, when cutoff $\Lambda$ tuning around 2 GeV, a loosely bound solution is obtained, and its RMS radius is around 1 fm with several MeV binding energy. The $\Xi_c^*\bar{K}$ and $\Xi_c\bar{K}^*$ systems are the dominant channels. If the value of cutoff around 2 GeV is a reasonable input, we may conclude that there exist a loosely bound state.

 \item For $I(J^P)=$ $0({1}/{2}^+)$ and $0({5}/{2}^{+})$, bound state solutions can be also obtained, by choosing the cutoff $\Lambda$ around 2 GeV or even larger. Moreover, the resulting sizes are too small around 0.5 fm or less.

 \item For $I(J^P)=$ $0({5}/{2}^-)$, as its cutoff $\Lambda$ is far away from 1 GeV, it cannot be a reasonable molecular candidate.

\end{itemize}

A possible reason for too small sizes for $I(J^P) =0(1/2^{\pm}, 5/2^{\pm})$ $\Xi_c^*\bar{K}/\Omega_c\eta/\Omega_c^*\eta/\Xi_c\bar{K}^*/\Xi_c'\bar{K}^*/\Omega_c\omega$ states is that they are dominated by higher mass channels such as $\Xi_c K^*$ and/or $\Xi_c^{\prime} K^*$. Measuring the binding energies of theses states from the higher mass thresholds, they turn to be around a few hundred MeV, which explains the small size of the bound state of the higher channels.

According to the above analysis, we have seen that there is one candidate of a molecular like state in the present approach, $\Xi_c^*\bar{K}/\Omega_c\eta/\Omega_c^*\eta/\Xi_c\bar{K}^*/\Xi_c'\bar{K}^*/\Omega_c\omega$ with $I(J^P)=0({3}/{2}^-)$, which is dominated by $\Xi_c^* \bar K$. In comparison with the analysis of a single $\Xi_c^* \bar K$ channel which does not accommodate a bound state solution with a cutoff $\Lambda \sim$ 1 GeV, it is implied that the coupled channel effect is very important in the $\Xi_c^*\bar{K}/\Omega_c\eta/\Omega_c^*\eta/\Xi_c\bar{K}^*/\Xi_c'\bar{K}^*/\Omega_c\omega$ interaction. %{\color{red}{In particular, when cutoff is taken as 1.37 GeV, one can reproduce the mass of newly reported $\Omega_c(3119)$ \cite{Aaij:2017nav}. Once the spin-parities of the these $\Omega_c$ states are further determined in future, the $\Omega_c$ state with $J^P=3/2^-$ is likely to explain as this $\Xi_c^*\bar{K}/\Omega_c\eta/\Omega_c^*\eta/\Xi_c\bar{K}^*/\Xi_c'\bar{K}^*/\Omega_c\omega$ state with $J^P=3/2^-$.}}

Besides the loose molecular state composed by the lowest channel, $\Xi_c^*\bar{K}$, we also obtain several bound solutions, where the dominant channels are $\Xi_c\bar{K}^*$ and/or $\Xi_c'\bar{K}^*$ channels. Their binding energies measured from these higher mass threshold $\tilde E$ reach several hundreds MeV, implying that the relevant OBE interaction is strongly attractive. In many of these cases, however, the cutoff $\Lambda$ is much larger than 1 GeV. By decreasing $\Lambda$ the interaction can be weakened, and these tightly bound states of the higher mass channels may become looser. As a result, they may appear as Feschbach type resonances. This is an interesting issue to be studied, which we would like to leave as a future study.

\section{Decay behavior for the predicted $\Omega_c$ state}\label{sec3}

\begin{figure}[!htbp]
\center
\includegraphics[width=4in]{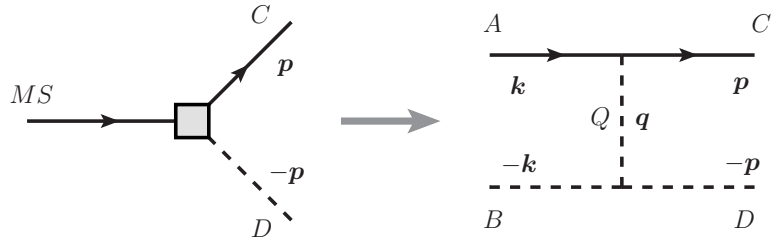}\\
\caption{Two body strong decay diagram for a molecular state $MS$ into a baryon $C$ and a meson $D$ (left). On the right hand, it is shown that $MS$ is composed by two colorless hadrons, a baryon $A$ and a meson $B$. $Q$ corresponds to the exchanged hadron. }\label{decay}
\end{figure}

In this section, we will discuss the two-body strong decay behaviors for the $\Xi_c^*\bar{K}/\Omega_c\eta/\Omega_c^*\eta/\Xi_c\bar{K}^*/\Xi_c'\bar{K}^*/\Omega_c\omega$ state with $I(J^P)=0({3}/{2}^-)$. A relevant diagram is presented in the left of the figure \ref{decay}.

In the rest frame of the molecular state $MS$ with mass $m_{MS}$, its two-body decay width for $MS\to C+D$ is
\begin{eqnarray}
d\Gamma &=& \frac{1}{2J+1}\frac{m_C|\bm{p}_{C}|}{8\pi^2 m_{MS}}|\mathcal{M}_I(MS\to C+D)|^2d\Omega,\\
|\bm{p}_C| &=& |\bm{p}_D|=\frac{1}{2m_{MS}}\left[\left(m_{MS}^2-(m_C+m_D)^2\right)\left(m_{MS}^2-(m_C-m_D)^2\right)\right]^{1/2},
\end{eqnarray}
where particles $C$ and $D$ are the final fermion and boson, respectively.

In figure \ref{decay}, the scattering amplitude for the process $MS\to C+D$ is related to that from process $A+B\to C+D$,
\begin{eqnarray}
\mathcal{M}_I (MS\to C+D) &=&\int d^3r\frac{d^3k}{(2\pi)^3}e^{-i\bm{k}\cdot\bm{r}}\psi(\bm{r})\frac{\mathcal{M}_I(A+B\to C+D)}{\sqrt{64m_Am_cm_{MS}}},
\end{eqnarray}
where $I$ stands for the isospin for $MS$ state, $\psi(\bm{r})$ is the wave function for the obtained $\Omega_c$ state, which has been obtained in the last section.

The allowed two-body decay channels for the $\Xi_c^*\bar{K}/\Omega_c\eta/\Omega_c^*\eta/\Xi_c\bar{K}^*/\Xi_c'\bar{K}^*/\Omega_c\omega$ state with $I(J^P)=0({3}/{2}^-)$ include $\Xi_c\bar{K}$ and $\Xi_c'\bar{K}$. The relevant scattering amplitude are written as
\begin{eqnarray}
\mathcal{M}_I^{(K^*)}(\Omega_c\eta\to \Xi_c\bar{K}) &=& i\frac{\lambda_Ig_V}{\sqrt{6}}\left(\varepsilon^{\mu\nu\lambda\kappa}-\varepsilon^{\mu\nu\kappa\lambda}\right)
              v_{\mu}\bar{u}_C\gamma^5\gamma_{\nu}q_{\lambda}u_A
              \frac{g_{\kappa\beta}-q_{\kappa}q_{\beta}/m_V^2}{q^2-m_V^2}\frac{g}{4}\left(P_{B\beta}+P_{D\beta}\right),\\
\mathcal{M}_I^{(K^*)}(\Omega_c\eta\to \Xi_c^{\prime}\bar{K}) &=& \left(\frac{\beta_Sg_V}{\sqrt{2}}\bar{u}_Cu_Av_{\nu}
              -\frac{\lambda_Sg_V}{3\sqrt{2}}\bar{u}_C(\gamma_{\mu}\gamma_{\nu}-\gamma_{\nu}\gamma_{\mu})q^{\mu}u_A\right)
              i\frac{g_{\nu\beta}-q_{\nu}q_{\beta}/m_V^2}{q^2-m_V^2}\frac{g}{4}\left(P_{B\beta}+P_{D\beta}\right),\\
\mathcal{M}_I^{(\rho,\omega,\phi, K^*)}(\Xi_c^*\bar{K}/\Omega_c^*\eta\to \Xi_c\bar{K}) &=& \frac{\lambda_Ig_V}{\sqrt{2}}\left(-\varepsilon^{\mu\nu\kappa\lambda}+\varepsilon^{\mu\nu\lambda\kappa}\right)v_{\mu}\bar{u}_{C}q_{\lambda}u_{A\nu}
            (-i)\frac{g^{\kappa\beta}-{q^{\kappa}q^{\beta}}/{m_V^2}}{q^2-m_V^2}
            \frac{g}{4}\left(P_{B\beta}+P_{D\beta}\right),\\
\mathcal{M}_I^{(\rho,\omega,\phi, K^*)}(\Xi_c^*\bar{K}/\Omega_c^*\eta\to \Xi_c'\bar{K}) &=& \left(\frac{\beta_Sg_V}{\sqrt{6}}\bar{u}_C\gamma^5(\gamma_{\mu}+v_{\mu})v_{\alpha}u_{A}^{\mu}-\frac{\lambda_Sg_V}{\sqrt{6}}\bar{u}_C\gamma^5q^{\mu}
              \left[(\gamma_{\mu}+v_{\mu})u_{A\alpha}-(\gamma_{\alpha}+v_{\alpha})u_{A\mu}\right]\right)\nonumber\\
              &&\times\frac{-g^{\alpha\beta}+q^{\alpha}q^{\beta}/m_V^2}{q^2-m_V^2}\frac{g}{4}(P_{B\beta}+P_{D\beta}),\\
\mathcal{M}_I^{(\rho,\omega,\phi)}(\Xi_c\bar{K}^*\to \Xi_c\bar{K}) &=& \frac{\beta_Bg_V}{\sqrt{2}}\bar{u}_Cu_Av_{\kappa}(-i)\frac{g^{\kappa\beta}-q^{\kappa}q^{\beta}/m_V^2}{q^2-m_V^2}
              \frac{g_{VVP}}{4}\varepsilon_{\mu\nu\alpha\beta}P_{B}^{\mu}\epsilon_B^{\nu}q^{\alpha},\\
\mathcal{M}_I^{(\rho,\omega,\phi)}(\Xi_c\bar{K}^*\to \Xi_c'\bar{K}) &=& \frac{\lambda_Ig_V}{\sqrt{6}}\left(\varepsilon^{\mu\nu\lambda\kappa}-\varepsilon^{\mu\nu\kappa\lambda}\right)
              v_{\mu}\bar{u}_C\gamma^5\gamma_{\nu}q_{\lambda}u_A
              \frac{g_{\kappa\beta}-q_{\kappa}q_{\beta}/m_V^2}{q^2-m_V^2}
              \frac{g_{VVP}}{4}\varepsilon_{\sigma\gamma\alpha\beta}P_B^{\sigma}\epsilon_B^{\gamma}q^{\alpha},\\
\mathcal{M}_{I}^{\pi}(\Xi_c\bar{K}^*\to \Xi_c'\bar{K}) &=& \frac{1}{\sqrt{3}}\frac{g_4}{f_{\pi}}\bar{u}_C\gamma^5(\gamma^{\mu}+v^{\mu})(q_{\mu})u_A\frac{1}{q^2-m_{\pi}^2}
              \frac{g}{4}\epsilon_B^{\nu}(q_{\nu}+P_{D\nu}),\\
\mathcal{M}_I^{(\rho,\omega,\phi,K^*)}(\Xi_c'\bar{K}^*/\Omega_c\omega\to \Xi_c\bar{K}) &=& \frac{\lambda_Ig_V}{\sqrt{6}}\left(\varepsilon^{\mu\nu\lambda\kappa}-\varepsilon^{\mu\nu\kappa\lambda}\right)
              v_{\mu}\bar{u}_C\gamma^5\gamma_{\nu}q_{\lambda}u_A
              \frac{g_{\kappa\beta}-q_{\kappa}q_{\beta}/m_V^2}{q^2-m_V^2}
              \frac{g_{VVP}}{4}\varepsilon_{\sigma\gamma\alpha\beta}P_B^{\sigma}\epsilon_B^{\gamma}q^{\alpha},\\
\mathcal{M}_I^{\pi,K}(\Xi_c'\bar{K}^*/\Omega_c\omega\to \Xi_c\bar{K}) &=& \frac{1}{\sqrt{3}}\frac{g_4}{f_{\pi}}\bar{u}_C\gamma^5(\gamma^{\mu}+v^{\mu})(q_{\mu})u_A\frac{1}{q^2-m_{\pi}^2}
              \frac{g}{4}\epsilon_B^{\nu}(q_{\nu}+P_{D\nu}),\\
\mathcal{M}_I^{(\rho,\omega,\phi,K^*)}(\Xi_c'\bar{K}^*/\Omega_c\omega\to \Xi_c'\bar{K}) &=& \left(\frac{\beta_Sg_V}{\sqrt{2}}\bar{u}_Cu_Av_{\nu}
              -\frac{\lambda_Sg_V}{3\sqrt{2}}\bar{u}_C(\gamma_{\mu}\gamma_{\nu}-\gamma_{\nu}\gamma_{\mu})q^{\mu}u_A\right)
              i\frac{g_{\nu\beta}-q_{\nu}q_{\beta}/m_V^2}{q^2-m_V^2}
              \frac{g_{VVP}}{4}\varepsilon_{\kappa\sigma\alpha\beta}P_B^{\kappa}\epsilon_B^{\sigma}q^{\alpha},\\
\mathcal{M}_I^{\pi,K}(\Xi_c'\bar{K}^*/\Omega_c\omega\to \Xi_c'\bar{K}) &=& \frac{g_1}{2f_{\pi}}\varepsilon^{\mu\nu\lambda\kappa}v_{\kappa}\bar{u}_C\gamma_{\mu}\gamma_{\lambda}q_{\nu}u_1\frac{i}{q^2-m_{\pi}^2}
             \frac{g}{4}\epsilon_B^{\alpha}(q_{\alpha}+P_{D\alpha}).
\end{eqnarray}

Using the wave function of the $\Xi_c^*\bar{K}/\Omega_c\eta/\Omega_c^*\eta/\Xi_c\bar{K}^*/\Xi_c'\bar{K}^*/\Omega_c\omega$ state with $I(J^P)=0({3}/{2}^-)$, the decay widths for all the allowed decay channels are collected in three groups, which are based on the choices of cutoff $\Lambda$ in table \ref{num1}.
\renewcommand\tabcolsep{0.6cm}
\renewcommand{\arraystretch}{1.6}
\begin{table}[!hbtp]
\caption{Allowed two-body decay with for the predicted $\Omega_c$ state.}\label{width}
\begin{tabular}{ccccc}
\toprule[1pt]
$\Lambda$ (GeV)      &$E$ (MeV)       &$\Gamma_{\Omega_c(3140)\to\Xi_c\bar{K}}$ (MeV)    &$\Gamma_{\Omega_c(3140)\to\Xi_c'\bar{K}}$ (MeV)      &$\Gamma_{\text{total}}$ (MeV)\\\hline
1.270                 &-0.58            &$\sim0 $       &0.060                                 &0.060\\
1.320                 &-6.00            &0.002          &1.909                                 &1.911\\
1.370                 &-20.55           &0.005          &16.360                                &16.365\\
\bottomrule[1pt]
\end{tabular}
\end{table}

In table \ref{width}, the total decay width for the $\Xi_c^*\bar{K}/\Omega_c\eta/\Omega_c^*\eta/\Xi_c\bar{K}^*/\Xi_c'\bar{K}^*/\Omega_c\omega$ state with $I(J^P)=0({3}/{2}^-)$ is around several MeV, and $\Xi'_c\bar{K}$ is the dominant decay channel. With the increasing of cutoff, the total decay width becomes larger and larger. For the $\Xi_c\bar{K}$ channel, the decay width is less than 1 MeV. A measurement in the $\Xi_c^\prime \bar K$ channel is helpful to further understand the structure of $\Omega_c$ states.

\section{Other partners}

After we discuss the isoscalar $\Xi_c^*\bar{K}/\Omega_c\eta/\Omega_c^*\eta/\Xi_c\bar{K}^*/\Xi_c'\bar{K}^*/\Omega_c\omega$ systems, in the following, we study their isovector partners by using the same model. The $\Xi_c^*\bar{K}$, $\Xi_c\bar{K}^*$, $\Xi_c'\bar{K}^*$, and $\Omega_c\rho$ channels are considered. The corresponding numerical results are presented in table \ref{num2}.

\renewcommand\tabcolsep{0.65cm}
\renewcommand{\arraystretch}{1.7}
\begin{table}[!hbtp]
\caption{Bound state preperties (binding energy $E$ and root-mean-square radius $r_{RMS}$) for all the investigated systems after the coupled channel effects are considered. Probability for the different channels are also given. $E$, $r_{RMS}$, and $\Lambda$ are in units of MeV, fm, and GeV, respectively. To emphasize the dominant channel, we label the probability for the corresponding channel in a bold manner.}\label{num2}
\begin{tabular}{c|ccc|cccccc}
\toprule[1pt]
 $I(J^{P})$      &$\Lambda$    &$E$       &$r_{RMS}$
      &$\Xi_{c}^*\bar{K}(\%)$       &$\Xi_{c}\bar{K}^*(\%)$     &$\Xi_{c}'\bar{K}^*(\%)$                &$\Omega_c\rho(\%)$   \\\midrule[1pt]
$1({1}/{2}^-)$           &1.656     &-0.96     &0.36       &0.61      &$\bm{46.95}$     &$\bm{50.48}$     &1.85\\
                         &1.660     &-5.06     &0.35       &0.60      &$\bm{46.82}$     &$\bm{50.72}$     &1.98\\
                         &1.664     &-9.21     &0.35       &0.60      &$\bm{46.45}$     &$\bm{50.94}$     &2.00\\
$1({1}/{2}^+)$           &1.820     &-0.46     &0.75       &8.31      &$\bm{55.20}$     &20.41     &16.07\\
                         &1.824     &-3.40     &0.63       &7.70      &$\bm{55.39}$     &20.46     &16.19\\
                         &1.828     &-6.36     &0.60       &7.36      &$\bm{55.65}$     &20.51     &16.46\\\hline
$1({3}/{2}^-)$           &1.460     &-0.67     &4.85       &$\bm{95.89}$     &3.60      &0.24      &0.27\\
                         &1.510     &-4.32     &2.37       &$\bm{90.00}$     &8.71      &0.53      &0.76\\
                         &1.560     &-11.35    &1.47       &$\bm{81.86}$     &13.93     &0.79      &1.42\\
$1({3}/{2}^+)$           &2.230     &-1.90     &0.48       &4.58      &23.55     &$\bm{48.32}$     &23.53\\
                         &2.232     &-4.26     &0.45       &4.43      &23.51     &$\bm{48.44}$     &23.61\\
                         &2.234     &-6.63     &0.44       &4.32      &23.45     &$\bm{48.53}$     &23.68\\\hline
$1({5}/{2}^-)$           &2.812     &-0.90     &0.40       &0.61      &20.58     &$\bm{56.55}$     &22.26\\
                         &2.816     &-5.34     &0.40       &0.61      &20.48     &$\bm{56.58}$     &22.31\\
                         &2.816     &-5.34     &0.40       &0.61      &20.40     &$\bm{55.60}$     &22.38\\
$1({5}/{2}^+)$           &3.590     &-1.27     &0.43       &0.92      &$\bm{37.30}$     &$\bm{26.38}$     &$\bm{35.39}$\\
                         &3.600     &-4.76     &0.42       &0.88      &$\bm{37.25}$     &$\bm{26.39}$     &$\bm{35.48}$\\
                         &3.610     &-8.28     &0.42       &0.85      &$\bm{37.19}$     &$\bm{26.39}$     &$\bm{35.56}$\\
\bottomrule[1pt]
\end{tabular}
\end{table}

Finally, we find that the $\Xi_c^*\bar{K}/\Xi_c\bar{K}^*/\Xi_c'\bar{K}^*/\Omega_c\rho$ state with $I(J^P)=1({3}/{2}^-)$ may be the other loosely bound molecular state, where the $\Xi_c^*\bar{K}$ channel is also dominant. For the $I(J^P)=$ $1({1}/{2}^{\pm})$, $1({3}/{2}^{+})$, and $1({5}/{2}^{\pm})$, because the value of the cutoff $\Lambda$ is away from 1 GeV and/or their RMS radii are very small, they are not possible loose molecular candidates.

As a byproduct, we further extend our study on the $\Xi_c^*K$ system. According to the G-parity rule \cite{Klempt:2002ap}, the properties of the $\omega$ and $\phi$ exchanges are exactly opposite to those in the $\Xi_c^*\bar{K}$ system. After solving the Schr\"{o}dinger equation, our results indicate that
\begin{itemize}
  \item For the isoscalar $\Xi_c^*K$ system, bound solutions can be obtained when cutoff is taken larger than 1.7 GeV. Thus, the interaction from considered one single channel is not strong enough to generate a loosely bound state.
  \item For the isovector $\Xi_c^*K$ system, there doesn't exist any binding solutions by tuning cutoff $\Lambda$ from 1 to 5 GeV.
\end{itemize}

When the $\Xi_c^*K/\Sigma_c\rho/\Sigma_c\omega/\Sigma_c^*\rho/\Sigma_c^*\omega/\Xi_cK^*/\Xi_c^{\prime}K^*$ coupled channel system is further considered, finally, we can predict two possible loosely bound molecular states, the isoscalar $\Xi_c^*K/\Sigma_c\rho/\Sigma_c^*\rho/\Xi_cK^*/\Xi_c^{\prime}K^*$ state with $J^P=3/2^-$ and the isovector $\Xi_c^*K/\Sigma_c\rho/\Sigma_c\omega/\Sigma_c^*\rho/\Sigma_c^*\omega/\Xi_cK^*/\Xi_c^{\prime}K^*$ state with $J^P=3/2^-$. For the isoscalar state, the $\Xi_c^*K$ channel is dominant with more than 80 percent probability. For the isovector state, the remaining channels $\Sigma_c\rho/\Sigma_c\omega/\Sigma_c^*\rho/\Sigma_c^*\omega/\Xi_cK^*/\Xi_c^{\prime}K^*$ play much more important roles, as their larger probabilities.

\section{Summary}\label{sec4}

Since the observation of the hidden-charm pentaquarks $P_c(4380)$ and $P_c(4450)$, searching for heavy pentaquarks becomes a very hot topic of hadron physics. In this work, by adopting the OBE model and coupled channel effect, we carry out a systematic investigation of possible $\Omega_c-$like molecular state from the $\Xi_c^*\bar{K}/\Omega_c\eta/\Omega_c^*\eta/\Xi_c\bar{K}^*/\Xi_c'\bar{K}^*/\Omega_c\omega$ interaction.

In order to justify the existence of a molecular state, we borrow the experience in the deuteron. Finally, we find a reasonable loose $\Omega_c-$like molecular state, the $\Xi_c^*\bar{K}/\Omega_c\eta/\Omega_c^*\eta/\Xi_c\bar{K}^*/\Xi_c'\bar{K}^*/\Omega_c\omega$ state with $I(J^P)=0(3/2^-)$. Compared to the probabilities for all the discussed channels, it is mainly made up by the S-wave $\Xi_c^*\bar{K}$ component. And then, we further study the two-body strong decay behavior of the $\Xi_c^*\bar{K}/\Omega_c\eta/\Omega_c^*\eta/\Xi_c\bar{K}^*/\Xi_c'\bar{K}^*/\Omega_c\omega$ state with $I(J^P)=0(3/2^-)$. Our results indicate that its two-body strong decay width is around several MeV, and the $\Xi_c^{\prime}$ decay channel is the dominant. {In comparison with the observation of the LHCb \cite{Aaij:2017nav}, the mass of this $\Xi_c^*\bar{K}/\Omega_c\eta/\Omega_c^*\eta/\Xi_c\bar{K}^*/\Xi_c'\bar{K}^*/\Omega_c\omega$ state with $I(J^P)=0(3/2^-)$ is very close to their threshold, especially the $\Omega_c(3090)$ and $\Omega_c(3119)$. If the spin-parities of the five $\Omega_c$ states is further determined in the further, the $\Omega_c$ with $I(J^P)=0(3/2^-)$ can be easily related to the $\Xi_c^*\bar{K}/\Omega_c\eta/\Omega_c^*\eta/\Xi_c\bar{K}^*/\Xi_c'\bar{K}^*/\Omega_c\omega$ state with $I(J^P)=0(3/2^-)$. Meanwhile, we may predict an isovector molecular partner, the $\Xi_c^*\bar{K}/\Xi_c\bar{K}^*/\Xi_c'\bar{K}^*/\Omega_c\rho$ state with $I(J^P)=1({3}/{2}^-)$.}

\section*{ACKNOWLEDGMENTS}

This project is partly supported by the National Natural Science Foundation of China under Grant No. 11222547, No. 11175073, and No. 11647301 and the Fundamental Research Funds for the Central Universities. Rui Chen is supported by the China Scholarship Council. Xiang Liu is also supported in part by the National Program for Support of Top-notch Young Professionals. Atsushi Hosaka is supported by the JSPS KAKENHI (the Grant-in-Aid for Scientific Research from Japan Society for the Promotion of Science (JSPS)) with Grant No. JP26400273(C).

\appendix

\section{Relevant subpotentials}\label{app01}
 The concrete subpotentials in equation (\ref{hh}) are expressed as
\begin{eqnarray}
\mathcal{V}_{I,J^P}^{\Xi_c^*\bar{K}\to\Xi_c^*\bar{K}}(r) &=& -\frac{1}{2}l_Sg_{\sigma}\mathcal{D}_1(J^P)Y(\Lambda,m_{\sigma},r)
                   -\frac{1}{16}\beta_Sg_Vg\mathcal{D}_1(J^P)\left[\mathcal{G}(I)Y(\Lambda,m_{\rho},r)+Y(\Lambda,m_{\omega},r)-2Y(\Lambda,m_{\phi},r)\right],\label{pot2}\\
\mathcal{V}_{I,J^P}^{\Xi_c\bar{K}^*\to\Xi_c^*\bar{K}}(r) &=&\frac{1}{8}\frac{g_4g}{f_{\pi}}\frac{1}{\sqrt{M_{\bar{K}}M_{\bar{K}^*}}}
                  \left[\frac{1}{3}\mathcal{E}_1(J^P)\nabla^2
                  +\frac{1}{3}\mathcal{F}_1(J^P)r\frac{\partial}{\partial r}\frac{1}{r}\frac{\partial}{\partial r}\right]\mathcal{G}(I)U(\Lambda_0,m_{\pi0},r)\nonumber\\
                  &&-\frac{1}{8}\frac{g_4g}{f_{\pi}}\frac{1}{\sqrt{M_{\bar{K}}M_{\bar{K}^*}}}
                  \left[\frac{1}{3}\mathcal{E}_1(J^P)\nabla^2
                  +\frac{1}{3}\mathcal{F}_1(J^P)r\frac{\partial}{\partial r}\frac{1}{r}\frac{\partial}{\partial r}\right]Y(\Lambda_0,m_{\eta0},r)\nonumber\\
                  &&-\frac{\lambda_Ig_Vg_{VVP}}{2\sqrt{2}}
                  \left[\frac{2}{3}\mathcal{E}_1(J^P)\nabla^2
                  -\frac{1}{3}\mathcal{F}_1(J^P)r\frac{\partial}{\partial r}\frac{1}{r}\frac{\partial}{\partial r}\right]
                  \left[\mathcal{G}(I)Y(\Lambda_0,m_{\rho0},r)+Y(\Lambda_0,m_{\omega0},r)-2Y(\Lambda_0,m_{\phi0},r)\right],\\
\mathcal{V}_{I,J^P}^{\Xi_c'\bar{K}^*\to\Xi_c^*\bar{K}}(r) &=& -\frac{\sqrt{3}}{16\sqrt{2}}\frac{g_1g}{f_{\pi}}\frac{1}{\sqrt{M_{\bar{K}}M_{\bar{K}^*}}}
                  \left[\frac{1}{3}\mathcal{E}_1(J^P)\nabla^2
                  +\frac{1}{3}\mathcal{F}_1(J^P)r\frac{\partial}{\partial r}\frac{1}{r}\frac{\partial}{\partial r}\right]\mathcal{G}(I)U(\Lambda_1,m_{\pi1},r)\nonumber\\
                  &&-\frac{\sqrt{3}}{48\sqrt{2}}\frac{g_1g}{f_{\pi}}\frac{1}{\sqrt{M_{\bar{K}}M_{\bar{K}^*}}}
                  \left[\frac{1}{3}\mathcal{E}_1(J^P)\nabla^2
                  +\frac{1}{3}\mathcal{F}_1(J^P)r\frac{\partial}{\partial r}\frac{1}{r}\frac{\partial}{\partial r}\right]Y(\Lambda_1,m_{\eta1},r)\nonumber\\
                  &&-\frac{\lambda_Sg_Vg_{VVP}}{8\sqrt{3}}
                  \left[\frac{2}{3}\mathcal{E}_1(J^P)\nabla^2
                  -\frac{1}{3}\mathcal{F}_1(J^P)r\frac{\partial}{\partial r}\frac{1}{r}\frac{\partial}{\partial r}\right]
                  \left[\mathcal{G}(I)Y(\Lambda_1,m_{\rho1},r)+Y(\Lambda_1,m_{\omega1},r)-2Y(\Lambda_1,m_{\phi1},r)\right],\\
\mathcal{V}_{I,J^P}^{\Xi_c\bar{K}^*\to\Xi_c\bar{K}^*}(r) &=& 2l_Bg_{\sigma}\mathcal{D}_{2}(J^P)Y(\Lambda,m_{\sigma},r)
                 +\frac{1}{8}\beta_Bg_Vg\mathcal{D}_2(J^P)\left[\mathcal{G}(I)Y(\Lambda,m_{\rho},r)+Y(\Lambda,m_{\omega},r)-2Y(\Lambda,m_{\phi},r)\right],\\
\mathcal{V}_{I,J^P}^{\Xi_c\bar{K}^*\to\Xi_c'\bar{K}^*}(r) &=&\frac{1}{4\sqrt{3}}\frac{g_4g_{VVP}}{f_{\pi}}
                  \left[\frac{1}{3}\mathcal{E}_2(J^P)\nabla^2
                  +\frac{1}{3}\mathcal{F}_2(J^P)r\frac{\partial}{\partial r}\frac{1}{r}\frac{\partial}{\partial r}\right]\mathcal{G}(I)Y(\Lambda_2,m_{\pi2},r)\nonumber\\
                  &&-\frac{1}{4\sqrt{3}}\frac{g_4g_{VVP}}{f_{\pi}}
                  \left[\frac{1}{3}\mathcal{E}_2(J^P)\nabla^2
                  +\frac{1}{3}\mathcal{F}_2(J^P)r\frac{\partial}{\partial r}\frac{1}{r}\frac{\partial}{\partial r}\right]Y(\Lambda_2,m_{\eta2},r)\nonumber\\
                  &&+\frac{1}{4\sqrt{6}}\lambda_Ig_Vg\frac{1}{M_{\bar{K}^*}}\mathcal{H}_1(J^P)\frac{1}{r}\frac{\partial}{\partial r}
                  \left[\mathcal{G}(I)Y(\Lambda_2,m_{\rho2},r)+Y(\Lambda_2,m_{\omega2},r)-2Y(\Lambda_2,m_{\phi2},r)\right]\nonumber\\
                  &&+\frac{\lambda_Ig_Vg}{4\sqrt{6}M_{\bar{K}^*}}
                  \left[\frac{2}{3}\mathcal{E}_2(J^P)\nabla^2
                  -\frac{1}{3}\mathcal{F}_2(J^P)r\frac{\partial}{\partial r}\frac{1}{r}\frac{\partial}{\partial r}\right]
                  \left[\mathcal{G}(I)Y(\Lambda_2,m_{\rho2},r)+Y(\Lambda_2,m_{\omega2},r)-2Y(\Lambda_2,m_{\phi2},r)\right],\\
\mathcal{V}_{I,J^P}^{\Xi_c'\bar{K}^*\to\Xi_c'\bar{K}^*}(r) &=& -\frac{1}{2}g_{\sigma}l_S\mathcal{D}_{2}(J^P)Y(\Lambda,m_{\sigma},r)
                  +\frac{\sqrt{3}}{8\sqrt{2}}\frac{g_1g_{VVP}}{f_{\pi}}
                  \left[\frac{1}{3}\mathcal{E}_2(J^P)\nabla^2
                  +\frac{1}{3}\mathcal{F}_2(J^P)r\frac{\partial}{\partial r}\frac{1}{r}\frac{\partial}{\partial r}\right]\mathcal{G}(I)Y(\Lambda,m_{\pi},r)\nonumber\\
                  &&+\frac{\sqrt{3}}{24\sqrt{2}}\frac{g_1g_{VVP}}{f_{\pi}}
                  \left[\frac{1}{3}\mathcal{E}_2(J^P)\nabla^2
                  +\frac{1}{3}\mathcal{F}_2(J^P)r\frac{\partial}{\partial r}\frac{1}{r}\frac{\partial}{\partial r}\right]Y(\Lambda,m_{\eta},r)\nonumber\\
                  &&-\frac{1}{16}\beta_Sg_Vg\mathcal{D}_{2}(J^P)\left[\mathcal{G}(I)Y(\Lambda,m_{\rho},r)+Y(\Lambda,m_{\omega},r)-2Y(\Lambda,m_{\phi},r)\right]\nonumber\\
                  &&-\frac{1}{48}\frac{\lambda_Sg_Vg}{M_{\Xi_c'}}\mathcal{D}_{2}(J^P)
                  \left[\mathcal{G}(I)\nabla^2Y(\Lambda,m_{\rho},r)+\nabla^2Y(\Lambda,m_{\omega},r)-2\nabla^2Y(\Lambda,m_{\phi},r)\right]\nonumber\\
                  &&+\frac{1}{24}\frac{\lambda_Sg_Vg}{M_{\bar{K}^*}}\mathcal{H}_1(J^P)\frac{1}{r}\frac{\partial}{\partial r}
                  \left[\mathcal{G}(I)Y(\Lambda,m_{\rho},r)+Y(\Lambda,m_{\omega},r)-2Y(\Lambda,m_{\phi},r)\right]\nonumber\\
                  &&-\frac{1}{24}\frac{\lambda_Sg_Vg}{M_{\bar{K}^*}}
                  \left[\frac{2}{3}\mathcal{E}_2(J^P)\nabla^2
                  -\frac{1}{3}\mathcal{F}_2(J^P)r\frac{\partial}{\partial r}\frac{1}{r}\frac{\partial}{\partial r}\right]
                  \left[\mathcal{G}(I)Y(\Lambda,m_{\rho},r)+Y(\Lambda,m_{\omega},r)-2Y(\Lambda,m_{\phi},r)\right],\label{pot3}\\
\mathcal{V}_{I,J^P}^{\Xi_c^*\bar{K}\to\Omega_c\eta}(r) &=& \frac{1}{8\sqrt{2}}\beta_Sg_Vg\mathcal{D}_3(J^P)Y(\Lambda_3,m_{K^*3},r)
                                    +\frac{1}{8\sqrt{2}}\frac{\lambda_Sg_Vg}{\sqrt{m_{\eta}m_K}}\mathcal{H}_2(J^P)\frac{1}{r}\frac{\partial}{\partial r}Y(\Lambda_3,m_{K^*3},r),\\
\mathcal{V}_{I,J^P}^{\Xi_c^*\bar{K}\to\Omega_c^*\eta}(r) &=& -\frac{1}{4\sqrt{6}}\beta_Sg_Vg\mathcal{D}_1(J^P)Y(\Lambda_4,m_{K^*4},r),\\
\mathcal{V}_{I,J^P}^{\Omega_c\omega\to\Xi_c^*\bar{K}}(r) &=& \frac{\sqrt{3}}{8}\frac{g_1g}{f_{\pi}}\frac{1}{\sqrt{m_{\omega}m_K}}\left[\frac{1}{3}\mathcal{E}_1(J^P)\nabla^2
                  +\frac{1}{3}\mathcal{F}_1(J^P)r\frac{\partial}{\partial r}\frac{1}{r}\frac{\partial}{\partial r}\right]Y(\Lambda_5,m_{K5},r)\nonumber\\
                  &&+\frac{\sqrt{6}}{12}\lambda_Sg_Vg_{VVP}\left[\frac{2}{3}\mathcal{E}_1(J^P)\nabla^2
                  -\frac{1}{3}\mathcal{F}_1(J^P)r\frac{\partial}{\partial r}\frac{1}{r}\frac{\partial}{\partial r}\right]Y(\Lambda_5,m_{K^*5},r),\\
\mathcal{V}_{I,J^P}^{\Omega_c\eta\to\Omega_c\eta}(r) &=& \frac{1}{6\sqrt{2}}\beta_Sg_VgY(\Lambda,m_{\phi},r)+\frac{1}{18\sqrt{2}}\frac{\lambda_Sg_Vg}{M_{\Omega_c}}\nabla^2Y(\Lambda,m_{\phi},r),\\
\mathcal{V}_{I,J^P}^{\Omega_c^*\eta\to\Omega_c\eta}(r) &=& -\frac{1}{6\sqrt{6}}\beta_Sg_Vg\mathcal{D}_2(J^P)Y(\Lambda_6,m_{\phi6},r)
                  -\frac{1}{6\sqrt{6}}\frac{\lambda_Sg_Vg}{m_{\eta}}\mathcal{H}_2(J^P)\frac{1}{r}\frac{\partial}{\partial r}Y(\Lambda_6,m_{\phi6},r),\\
\mathcal{V}_{I,J^P}^{\Omega_c\omega\to\Omega_c\eta}(r) &=& -\frac{1}{12\sqrt{3}}\frac{g_1g}{f_{\pi}}\left[\frac{1}{3}\mathcal{E}_3(J^P)\nabla^2
                  +\frac{1}{3}\mathcal{F}_3(J^P)r\frac{\partial}{\partial r}\frac{1}{r}\frac{\partial}{\partial r}\right]Y(\Lambda_7,m_{\eta7},r),\\
\mathcal{V}_{I,J^P}^{\Xi_c\bar{K}^*\to\Omega_c\eta}(r) &=& \frac{1}{6\sqrt{2}}\frac{g_4g}{f_{\pi}}\frac{1}{\sqrt{m_{\eta}m_{K^*}}}\left[\frac{1}{3}\mathcal{E}_3(J^P)\nabla^2
                  +\frac{1}{3}\mathcal{F}_3(J^P)r\frac{\partial}{\partial r}\frac{1}{r}\frac{\partial}{\partial r}\right]Y(\Lambda_8,m_{K8},r)\nonumber\\
                  &&+\frac{1}{12}\lambda_Ig_Vg_{VVP}\sqrt{\frac{m_{K^*}}{m_{\eta}}}\left[\frac{2}{3}\mathcal{E}_3(J^P)\nabla^2
                  -\frac{1}{3}\mathcal{F}_3(J^P)r\frac{\partial}{\partial r}\frac{1}{r}\frac{\partial}{\partial r}\right]Y(\Lambda_8,m_{K^*8},r),\\
\mathcal{V}_{I,J^P}^{\Xi_c'\bar{K}^*\to\Omega_c\eta}(r) &=& -\frac{1}{4\sqrt{3}}\frac{g_1g}{f_{\pi}}\frac{1}{\sqrt{m_{\eta}m_{K^*}}}\left[\frac{1}{3}\mathcal{E}_3(J^P)\nabla^2
                  +\frac{1}{3}\mathcal{F}_3(J^P)r\frac{\partial}{\partial r}\frac{1}{r}\frac{\partial}{\partial r}\right]Y(\Lambda_9,m_{K9},r)\nonumber\\
                  &&+\frac{1}{4\sqrt{6}}\frac{\beta_Sg_Vg_{VVP}}{\sqrt{m_{\eta}m_{K^*}}}\mathcal{H}_3(J^P)\frac{1}{r}\frac{\partial}{\partial r}Y(\Lambda_9,m_{K^*9},r)\nonumber\\
                  &&+\frac{1}{12\sqrt{6}}\lambda_Sg_Vg_{VVP}\sqrt{\frac{m_{K^*}}{m_{\eta}}}\left[\frac{2}{3}\mathcal{E}_3(J^P)\nabla^2
                  -\frac{1}{3}\mathcal{F}_3(J^P)r\frac{\partial}{\partial r}\frac{1}{r}\frac{\partial}{\partial r}\right]Y(\Lambda_9,m_{K^*9},r),\\
\mathcal{V}_{I,J^P}^{\Omega_c^*\eta\to\Omega_c^*\eta}(r) &=& \frac{1}{6\sqrt{2}}\beta_Sg_Vg\mathcal{D}_1(J^P)Y(\Lambda,m_{\phi},r),\\
\mathcal{V}_{I,J^P}^{\Xi_c\bar{K}^*\to\Omega_c^*\eta}(r) &=& \sqrt{\frac{2}{3}}\frac{g_4g}{f_{\pi}}\frac{1}{\sqrt{m_{\eta}m_{K^*}}}\left[\frac{1}{3}\mathcal{E}_1(J^P)\nabla^2
                  +\frac{1}{3}\mathcal{F}_1(J^P)r\frac{\partial}{\partial r}\frac{1}{r}\frac{\partial}{\partial r}\right]Y(\Lambda_{10},m_{K{10}},r)\nonumber\\
                  &&-\frac{1}{\sqrt{3}}\lambda_Ig_Vg_{VVP}\left[\frac{2}{3}\mathcal{E}_1(J^P)\nabla^2
                  -\frac{1}{3}\mathcal{F}_1(J^P)r\frac{\partial}{\partial r}\frac{1}{r}\frac{\partial}{\partial r}\right]Y(\Lambda_{10},m_{K^*{10}},r),\\
\mathcal{V}_{I,J^P}^{\Xi_c'\bar{K}^*\to\Omega_c^*\eta}(r) &=& -\frac{1}{8}\frac{g_1g}{f_{\pi}}\frac{1}{\sqrt{m_{\eta}m_{K^*}}}\left[\frac{1}{3}\mathcal{E}_1(J^P)\nabla^2
                  +\frac{1}{3}\mathcal{F}_1(J^P)r\frac{\partial}{\partial r}\frac{1}{r}\frac{\partial}{\partial r}\right]Y(\Lambda_{11},m_{K11},r)\nonumber\\
                  &&-\frac{1}{12}\lambda_Sg_Vg_{VVP}\left[\frac{2}{3}\mathcal{E}_1(J^P)\nabla^2
                  -\frac{1}{3}\mathcal{F}_1(J^P)r\frac{\partial}{\partial r}\frac{1}{r}\frac{\partial}{\partial r}\right]Y(\Lambda_{11},m_{K^*11},r),\\
\mathcal{V}_{I,J^P}^{\Omega_c\omega\to\Omega_c\omega}(r) &=& \frac{1}{4\sqrt{3}}\frac{g_1g_{VVP}}{f_{\pi}}\left[\frac{1}{3}\mathcal{E}_2(J^P)\nabla^2
                  +\frac{1}{3}\mathcal{F}_2(J^P)r\frac{\partial}{\partial r}\frac{1}{r}\frac{\partial}{\partial r}\right]Y(\Lambda,m_{\eta},r),\\
\mathcal{V}_{I,J^P}^{\Xi_c\bar{K}^*\to\Omega_c\omega}(r) &=& -\frac{1}{\sqrt{6}}\frac{g_4g_{VVP}}{f_{\pi}}\left[\frac{1}{3}\mathcal{E}_2(J^P)\nabla^2
                  +\frac{1}{3}\mathcal{F}_2(J^P)r\frac{\partial}{\partial r}\frac{1}{r}\frac{\partial}{\partial r}\right]Y(\Lambda_{12},m_{K12},r)\nonumber\\
                  &&+\frac{1}{2\sqrt{3}}\frac{\lambda_Ig_Vg}{\sqrt{m_{K^*}m_{\omega}}}\mathcal{H}_1(J^P)\frac{1}{r}\frac{\partial}{\partial r}Y(\Lambda_{12},m_{K^*12},r)\nonumber\\
                  &&+\frac{1}{2\sqrt{3}}\frac{\lambda_Ig_Vg}{\sqrt{m_{K^*}m_{\omega}}}\left[\frac{2}{3}\mathcal{E}_2(J^P)\nabla^2
                  -\frac{1}{3}\mathcal{F}_2(J^P)r\frac{\partial}{\partial r}\frac{1}{r}\frac{\partial}{\partial r}\right]Y(\Lambda_{12},m_{K^*12},r),\\
\mathcal{V}_{I,J^P}^{\Xi_c'\bar{K}^*\to\Omega_c\omega}(r) &=& -\frac{\sqrt{3}}{4}\frac{g_1g_{VVP}}{f_{\pi}}\left[\frac{1}{3}\mathcal{E}_2(J^P)\nabla^2
                  +\frac{1}{3}\mathcal{F}_2(J^P)r\frac{\partial}{\partial r}\frac{1}{r}\frac{\partial}{\partial r}\right]Y(\Lambda_{13},m_{K13},r)\nonumber\\
                  &&+\frac{1}{4\sqrt{2}}\beta_Sg_Vg\mathcal{D}_2(J^P)Y(\Lambda_{12},m_{K^*12},r)+\frac{1}{12\sqrt{2}}\lambda_Sg_Vg\mathcal{D}_2(J^P)\nabla^2Y(\Lambda_{13},m_{K^*13},r)\nonumber\\
                  &&-\frac{1}{6\sqrt{2}}\frac{\lambda_Sg_Vg}{\sqrt{m_{\omega}m_{K^*}}}\mathcal{H}_1(J^P)\frac{1}{r}\frac{\partial}{\partial r}Y(\Lambda_{12},m_{K^*13},r)\nonumber\\
                  &&+\frac{1}{6\sqrt{2}}\frac{\lambda_Sg_Vg}{\sqrt{m_{\omega}m_{K^*}}}\left[\frac{2}{3}\mathcal{E}_2(J^P)\nabla^2
                  -\frac{1}{3}\mathcal{F}_2(J^P)r\frac{\partial}{\partial r}\frac{1}{r}\frac{\partial}{\partial r}\right]Y(\Lambda_{13},m_{K^*13},r).\label{pot9}
\end{eqnarray}
In equation (\ref{pot2}), the function $U(\Lambda_i,m_{\pi i},{r})$ is defined as
\begin{eqnarray*}
U(\Lambda_i,m_{\pi i},{r}) &=&\frac{1}{4\pi r}(\cos(m_{\pi i}r)-e^{-\Lambda_i r})-\frac{\Lambda_i^2+m_{\pi i}^2}{8\pi \Lambda_i}e^{-\Lambda_i r},\quad\quad\quad i=0,1,2.\label{uu}
\end{eqnarray*}
The variables in Eqs. (\ref{pot2})-(\ref{pot9}) are
%\renewcommand\tabcolsep{0.18cm}
%\renewcommand{\arraystretch}{1.7}
%\begin{table}[!htbp]\center
%  \caption{The possible channels involved in our calculation. Here, the first column is the spin-parity quantum numbers corresponding to the channels.}\label{q0}
%  {\begin{tabular}{cllllll}
%\toprule[1pt]
%\\
%\bottomrule[1pt]
%\end{tabular}}
%\end{table}

{{\begin{eqnarray}
\left.\begin{array}{llll}
q_0 = \frac{M_{\Xi_c^*}^2+M_{\bar{K}^*}^2-M_{\Xi_c}^2-M_{K}^2}{2(M_{\Xi_c^*}+M_{\bar{K}})},\quad\quad\quad
&q_1 = \frac{M_{\Xi_c^*}^2+M_{\bar{K}^*}^2-M_{\Xi_c'}^2-M_{K}^2}{2(M_{\Xi_c^*}+M_{\bar{K}})},\quad\quad\quad
&q_2 = \frac{M_{\Xi_c^{\prime}}^2-M_{\Xi_c}^2}{2(M_{\Xi_c}+M_{\bar{K}^*}^2)},\quad\quad\quad
&q_3 = \frac{M_{\Xi_c^*}^2-M_{K}^2+m_{\eta}^2-M_{\Omega_c}^2}{2(M_{\Xi_c^*}+M_{\bar{K}})},\\
q_4 = \frac{M_{\Xi_c^*}^2-M_{K}^2+m_{\eta}^2-M_{\Omega_c^*}^2}{2(M_{\Xi_c^*}+M_{\bar{K}})},
&q_5 = \frac{M_{\Xi_c^*}^2-M_{K}^2+m_{\omega}^2-M_{\Omega_c}^2}{2(M_{\Xi_c^*}+M_{\bar{K}})},
&q_6 =\frac{-M_{\Omega_c}^2+M_{\Omega_c^*}^2}{2(M_{\Omega_c}+M_{\eta})},\quad\quad\quad
&q_7 =\frac{-M_{\eta}^2+M_{\omega}^2}{2(M_{\Omega_c}+M_{\eta})},\\
q_8 = \frac{M_{\Omega_c}^2-M_{\eta}^2+M_{K^*}^2-M_{\Xi_c}^2}{2(M_{\Omega_c}+M_{\eta})},
&q_9 = \frac{M_{\Omega_c}^2-M_{\eta}^2+M_{K^*}^2-M_{\Xi_c^{\prime}}^2}{2(M_{\Omega_c}+M_{\eta})},
&q_{10} = \frac{M_{\Omega_c^*}^2-M_{\eta}^2+M_{K^*}^2-M_{\Xi_c}^2}{2(M_{\Omega_c^*}+M_{\eta})},
&q_{11} =\frac{M_{\Omega_c^*}^2-M_{\eta}^2+M_{K^*}^2-M_{\Xi_c^{\prime}}^2}{2(M_{\Omega_c^*}+M_{\eta})},\\
q_{12} = \frac{M_{\Omega_c}^2-M_{\omega}^2+M_{K^*}^2-M_{\Xi_c}^2}{2(M_{\Omega_c}+M_{\omega})},
&q_{13} =  \frac{M_{\Omega_c}^2-M_{\omega}^2+M_{K^*}^2-M_{\Xi_c^{\prime}}^2}{2(M_{\Omega_c}+M_{\omega})},
&\Lambda_i^2= \Lambda^2-q_i^2,
&mE_i^2 =\left|mE^2-q_i^2\right|
.\end{array}\right.
\end{eqnarray}}}

In addition, $\mathcal{D}_i(J^P)$, $\mathcal{E}_{i}(J^P)$, $\mathcal{F}_i(J^P)$ and $\mathcal{H}_i(J^P)$ in Eqs. (\ref{pot2})-(\ref{pot9}), which are the spin-spin, tensor, and spin-orbit operators, respectively. For example,
\begin{eqnarray}\label{op}
\left.\begin{array}{ll}
\mathcal{D}_1=\sum_{a,b}^{m,n}C_{1/2,a;1,b}^{3/2,a+b}C_{1/2,m;1,n}^{3/2,m+n}\chi_3^{a\dag}\chi_1^m\bm{\epsilon}_1^n\cdot\bm{\epsilon}_3^{b\dag}, \quad\quad\quad\quad\quad\quad
&\mathcal{D}_2=\chi_3^{\dag}\chi_1\bm{\epsilon}_2\cdot\bm{\epsilon}_4^{\dag},\\
\mathcal{D}_3=\sum_{a,b}C_{1/2,a;1,b}^{3/2,a+b}\chi_3^{\dag}\left(\bm{\sigma}\cdot\bm{\epsilon}_{1b}\right)\chi_{1a},
&\mathcal{E}_1=\sum_{m,n}C_{1/2,m;1,n}^{3/2,m+n}\chi_3^{m\dag}\bm{\epsilon}_2\cdot\left(i\bm{\epsilon}_3^{n\dag}\times\bm{\sigma}\right)\chi_1,\\
\mathcal{E}_2=\chi_3^{\dag}\bm{\sigma}\cdot\left(i\bm{\epsilon}_2\times\bm{\epsilon}_4^{\dag}\right)\chi_1,
&\mathcal{E}_3=\chi_3^{\dag}(\bm{\sigma}\cdot\bm{\epsilon}_2)\chi_1,\\
\mathcal{F}_1=\sum_{m,n}C_{1/2,m;1,n}^{3/2,m+n}\chi_3^{m\dag}S\left(\hat{r},\bm{\epsilon}_2,i\bm{\epsilon}_3^{n\dag}\times\bm{\sigma}\right)\chi_1,
&\mathcal{F}_2=\chi_3^{\dag}S\left(\hat{r},\bm{\sigma},i\bm{\epsilon}_2\times\bm{\epsilon}_4^{\dag}\right)\chi_1,\\
\mathcal{F}_3=\chi_3^{\dag}S(\hat{r},\bm{\sigma},\bm{\epsilon}_2)\chi_1,
&\mathcal{H}_1=\chi_3^{\dag}\left(\bm{\epsilon}_2\cdot\bm{\epsilon}_4^{\dag}\right)\left(\bm{\sigma}\cdot\bm{L}\right)\chi_1,\\
\mathcal{H}_2=\sum_{a,b}C_{1/2,a;1,b}^{3/2,a+b}\chi_3^{\dag}(-i\bm{\sigma}\times\bm{\epsilon}_1^b)\cdot\bm{L}\chi_{1a},
&\mathcal{H}_3=\chi_3^{\dag}(\bm{L}\cdot\bm{\epsilon}_2)\chi_1,\\
S(\hat{r},\bm{a},\bm{b})=3(\hat{r}\cdot\bm{a})(\hat{r}\cdot\bm{b})-\bm{a}\cdot\bm{b}.
\end{array}\right.
\end{eqnarray}
The subscripts $(1,2,3,4)$ labeled in the polarization vector $\bm{\epsilon}$ and spin wave function $\chi$ correspond to the hadrons in the process $B(1){M}(2)\to B(3){M}(4)$, where $B$ and ${M}$ stand for the baryon and meson, respectively. When performing the numerical calculations, these corresponding operators in equation (\ref{op}) will be replaced by numerical matrixes, which are summarized in table \ref{matrix}.

\renewcommand\tabcolsep{0.4cm}
\renewcommand{\arraystretch}{1.5}
\begin{table}[!htbp]
  \caption{Non-zero matrix elements $\langle f|\mathcal{A}|i\rangle$ in various channels for the operators $\mathcal{A}$ in equation (\ref{op}).  \label{matrix}}
  \begin{tabular}{c|l|l|l|l|l|l}\toprule[1pt]
   {{{$\mathcal{A}$}}}
                    &$1/2^-$     &$1/2^+$     &$3/2^-$     &$3/2^+$    &$5/2^-$    &$5/2^+$\\\midrule[1pt]
   $\mathcal{D}_1$
            &$\left(\begin{array}{c}1\end{array}\right)$
                 &$\left(\begin{array}{c}1\end{array}\right)$
                      &$\left(\begin{array}{cc} 1 & 0 \\ 0 & 1\end{array}\right)$
                           &$\left(\begin{array}{c}1\end{array}\right)$
                                &$\left(\begin{array}{c}1\end{array}\right)$
                                     &$\left(\begin{array}{c}1\end{array}\right)$\\
                                            &&&&&\\
   $\mathcal{D}_2$
            &$\left(\begin{array}{cc} 1 & 0 \\ 0 & 1\end{array}\right)$
                 &$\left(\begin{array}{cc} 1 & 0 \\ 0 & 1\end{array}\right)$
                      &{$\left(\begin{array}{ccc} 1 & 0 & 0 \\ 0 & 1 & 0 \\ 0 & 0 & 1\end{array}\right)$}
                           &{$\left(\begin{array}{cc} 1 & 0 \\ 0 & 1\end{array}\right)$}
                                &{$\left(\begin{array}{cc} 1 & 0 \\ 0 & 1\end{array}\right)$}
                                     &{$\left(\begin{array}{c} 1\end{array}\right)$}\\
                                            &&&&&\\
   $\mathcal{D}_3$
             &$\left(\begin{array}{c} 0\end{array}\right)$
                  &$\left(\begin{array}{c} 0\end{array}\right)$
                        &$\left(\begin{array}{cc} 0    &\frac{2}{5 \sqrt{3}}\end{array}\right)$
                             &$\left(\begin{array}{c} -\frac{2}{\sqrt{15}}\end{array}\right)$
                                     &$\left(\begin{array}{c} -2 \sqrt{\frac{2}{21}}\end{array}\right)$
                                            &$\left(\begin{array}{c} 0\end{array}\right)$\\
                                            &&&&&\\
   $\mathcal{E}_1$
            &$\left(\begin{array}{cc} 0 & 1\end{array}\right)$
                 &$\left(\begin{array}{cc} 0 & 1\end{array}\right)$
                      &$\left(\begin{array}{ccc} 1 & 0 & 0 \\ 0 & 0 & 1\end{array}\right)$
                           &$\left(\begin{array}{ccc} 0 & 1\end{array}\right)$
                                &$\left(\begin{array}{cc}0 & 1\end{array}\right)$
                                     &$\left(\begin{array}{c}1\end{array}\right)$\\
                                            &&&&&\\
   $\mathcal{E}_2$
            &$\left(\begin{array}{cc} -2 & 0 \\ 0 & 1\end{array}\right)$
                 &$\left(\begin{array}{cc} -2 & 0 \\ 0 & 1\end{array}\right)$
                      &{$\left(\begin{array}{ccc} 1 & 0 & 0 \\ 0 & -2 & 0 \\ 0 & 0 & 1\end{array}\right)$}
                           &{$\left(\begin{array}{cc} -2 & 0 \\ 0 & 1\end{array}\right)$}
                                &{$\left(\begin{array}{cc} -2 & 0 \\ 0 & 1\end{array}\right)$}
                                     &{$\left(\begin{array}{c} 1\end{array}\right)$}\\
                                            &&&&&\\
   $\mathcal{E}_3$
             &$\left(\begin{array}{cc} \sqrt{3}    &0\end{array}\right)$
                   &$\left(\begin{array}{cc} \sqrt{3}    &0\end{array}\right)$
                        &$\left(\begin{array}{ccc} 0   &\sqrt{3}    &0\end{array}\right)$
                             &$\left(\begin{array}{cc} \sqrt{3}    &0\end{array}\right)$
                                     &$\left(\begin{array}{cc} \sqrt{3}    &0\end{array}\right)$\\
                                            &&&&&\\
   $\mathcal{F}_1$
            &$\left(\begin{array}{cc} -\sqrt{2} & 1\end{array}\right)$
                 &$\left(\begin{array}{cc} -\sqrt{2} & 1\end{array}\right)$
                      &$\left(\begin{array}{ccccc} 0 & 1 & -1 \\ -1 & -1 & 0\end{array}\right)$
                           &$\left(\begin{array}{cc}\frac{1}{\sqrt{5}} & -\frac{4}{5}\end{array}\right)$
                                &$\left(\begin{array}{cc}\sqrt{\frac{2}{7}} & -\frac{5}{7}\end{array}\right)$
                                     &$\left(\begin{array}{c}\frac{1}{5}\end{array}\right)$
                                     \\
                                            &&&&&\\
   $\mathcal{F}_2$
            &$\left(\begin{array}{cc} 0 & -\sqrt{2} \\ -\sqrt{2} & -2\end{array}\right)$
                 &$\left(\begin{array}{cc} 0 & -\sqrt{2} \\ -\sqrt{2} & -2\end{array}\right)$
                      &{$\left(\begin{array}{ccc} 0 & 1 & 2 \\ 1 & 0 & -1 \\ 2 & -1 & 0\end{array}\right)$}
                           &{$\left(\begin{array}{cc} 0 & \frac{1}{\sqrt{5}} \\ \frac{1}{\sqrt{5}} & \frac{8}{5}\end{array}\right)$}
                                &{$\left(\begin{array}{cc} 0 & \sqrt{\frac{2}{7}} \\ \sqrt{\frac{2}{7}} & \frac{10}{7}\end{array}\right)$}
                                     &{$\left(\begin{array}{c} -\frac{2}{5}\end{array}\right)$}\\
                                            &&&&&\\
   $\mathcal{F}_3$
             &$\left(\begin{array}{cc} 0     &-\sqrt{6}  \end{array}\right)$
                   &$\left(\begin{array}{cc} 0     &-\sqrt{6}  \end{array}\right)$
                        &$\left(\begin{array}{ccc} \sqrt{3}   &0     &-\sqrt{3}  \end{array}\right)$
                              &$\left(\begin{array}{cc} 0     &\sqrt{\frac{3}{5}}  \end{array}\right)$
                                     &$\left(\begin{array}{cc} 0     &\sqrt{\frac{6}{7}}  \end{array}\right)$\\
                                            &&&&&\\
   $\mathcal{H}_1$
             &$\left(\begin{array}{cc} 0 & 0 \\ 0 & -3\end{array}\right)$
                   &$\left(\begin{array}{cc} \frac{2}{3} & -\frac{2}{\sqrt{3}} \\ -\frac{2}{\sqrt{3}} & -\frac{5}{3}\end{array}\right)$
                        &{$\left(\begin{array}{ccc} 0 & 0 & 0 \\ 0 & 1 & -2 \\ 0 & -2 & -2\end{array}\right)$}
                              &$\left(\begin{array}{cc} -\frac{1}{3} & -\frac{2\sqrt{5}}{3} \\ -\frac{2\sqrt{5}}{3} & -\frac{2}{3}\end{array}\right)$
                                     &$\left(\begin{array}{cc} -\frac{2}{3} & -\frac{2\sqrt{14}}{3} \\ -\frac{2\sqrt{14}}{3} & -\frac{1}{3}\end{array}\right)$
                                            &{$\left(\begin{array}{c} 1\end{array}\right)$}\\
                                            &&&&&\\
   $\mathcal{H}_2$
             &$\left(\begin{array}{c} 0\end{array}\right)$
                  &$\left(\begin{array}{c} 0\end{array}\right)$
                        &$\left(\begin{array}{cc} 0    &\sqrt{3}\end{array}\right)$
                             &$\left(\begin{array}{c} \sqrt{\frac{5}{3}}\end{array}\right)$
                                     &$\left(\begin{array}{c} \sqrt{\frac{14}{3}}\end{array}\right)$
                                            &$\left(\begin{array}{c} 0   \end{array}\right)$\\
                                            &&&&&\\
   $\mathcal{H}_3$
             &$\left(\begin{array}{cc} 0     &0  \end{array}\right)$
                  &$\left(\begin{array}{cc} -\frac{2}{\sqrt{3}}     &-\sqrt{\frac{2}{3}}  \end{array}\right)$
                        &$\left(\begin{array}{ccc}  0     &-\sqrt{3}     &-\sqrt{3}  \end{array}\right)$
                             &$\left(\begin{array}{cc} \frac{1}{\sqrt{3}}     &-\sqrt{\frac{5}{3}}  \end{array}\right)$
                                     &$\left(\begin{array}{cc} \frac{2}{\sqrt{3}}     &-\sqrt{\frac{14}{3}}  \end{array}\right)$\\
   \bottomrule[1pt]
  \end{tabular}
\end{table}


\begin{thebibliography}{99}
%\cite{Chen:2016qju}
\bibitem{Chen:2016qju}
  H.~X.~Chen, W.~Chen, X.~Liu and S.~L.~Zhu,
  The hidden-charm pentaquark and tetraquark states,
  \href{http://www.sciencedirect.com/science/article/pii/S037015731630103X?via\%3Dihub}{Phys.\ Rept.\  {\bf 639}, 1 (2016)},
  %doi:10.1016/j.physrep.2016.05.004
  \href{https://arxiv.org/abs/1601.02092}{[arXiv:1601.02092 [hep-ph]]}.
  %%CITATION = doi:10.1016/j.physrep.2016.05.004;%%
  %113 citations counted in INSPIRE as of 06 Jan 2017

%\cite{Liu:2013waa}
\bibitem{Liu:2013waa}
  X.~Liu,
  An overview of $XYZ$ new particles,
  \href{https://link.springer.com/article/10.1007\%2Fs11434-014-0407-2}{Chin.\ Sci.\ Bull.\  {\bf 59}, 3815 (2014)},
  %doi:10.1007/s11434-014-0407-2
  \href{https://arxiv.org/abs/1312.7408}{[arXiv:1312.7408 [hep-ph]]}.
  %%CITATION = doi:10.1007/s11434-014-0407-2;%%
  %59 citations counted in INSPIRE as of 06 Jan 2017

%\cite{Aaij:2015tga}
\bibitem{Aaij:2015tga}
  R.~Aaij {\it et al.} [LHCb Collaboration],
  Observation of $J/\psi$ Resonances Consistent with Pentaquark States in $\Lambda_b^0\rightarrow J/\psi K^-p$ Decays,
  \href{https://journals.aps.org/prl/abstract/10.1103/PhysRevLett.115.072001}{Phys.\ Rev.\ Lett.\  {\bf 115}, 072001 (2015)},
  %doi:10.1103/PhysRevLett.115.072001
  \href{https://arxiv.org/abs/1507.03414}{[arXiv:1507.03414 [hep-ex]]}.
  %%CITATION = doi:10.1103/PhysRevLett.115.072001;%%
  %83 citations counted in INSPIRE as of 24 Nov 2015

%\cite{Chen:2015loa}
\bibitem{Chen:2015loa}
  R.~Chen, X.~Liu, X.~Q.~Li and S.~L.~Zhu,
  Identifying exotic hidden-charm pentaquarks,
  \href{https://journals.aps.org/prl/abstract/10.1103/PhysRevLett.115.132002}{Phys.\ Rev.\ Lett.\  {\bf 115}, 132002 (2015)},
  %doi:10.1103/PhysRevLett.115.132002
  \href{https://arxiv.org/abs/1507.03704}{[arXiv:1507.03704 [hep-ph]]}.
  %%CITATION = doi:10.1103/PhysRevLett.115.132002;%%
  %26 citations counted in INSPIRE as of 24 Nov 2015

%\cite{Chen:2015moa}
\bibitem{Chen:2015moa}
  H.~X.~Chen, W.~Chen, X.~Liu, T.~G.~Steele and S.~L.~Zhu,
  Towards exotic hidden-charm pentaquarks in QCD,
  \href{https://journals.aps.org/prl/abstract/10.1103/PhysRevLett.115.172001}{Phys.\ Rev.\ Lett.\  {\bf 115}, 172001 (2015)},
  %doi:10.1103/PhysRevLett.115.172001
  \href{https://arxiv.org/abs/1507.03717}{[arXiv:1507.03717 [hep-ph]]}.
  %%CITATION = doi:10.1103/PhysRevLett.115.172001;%%
  %22 citations counted in INSPIRE as of 29 Nov 2015

%\cite{Roca:2015dva}
\bibitem{Roca:2015dva}
  L.~Roca, J.~Nieves and E.~Oset,
  LHCb pentaquark as a $\bar{D}^*\Sigma_c-\bar{D}^*\Sigma_c^*$ molecular state,
  \href{https://journals.aps.org/prd/abstract/10.1103/PhysRevD.92.094003}{Phys.\ Rev.\ D {\bf 92}, 094003 (2015)},
  %doi:10.1103/PhysRevD.92.094003
  \href{https://arxiv.org/abs/1507.04249}{[arXiv:1507.04249 [hep-ph]]}.
  %%CITATION = doi:10.1103/PhysRevD.92.094003;%%
  %26 citations counted in INSPIRE as of 29 Nov 2015

%\cite{He:2015cea}
\bibitem{He:2015cea}
  J.~He,
  $\bar{D}\Sigma^*_c$ and $\bar{D}^*\Sigma_c$ interactions and the LHCb hidden-charmed pentaquarks,
  \href{http://www.sciencedirect.com/science/article/pii/S0370269315010199?via\%3Dihub}{Phys.\ Lett.\ B {\bf 753}, 547 (2016)},
%  doi:10.1016/j.physletb.2015.12.071
  \href{https://arxiv.org/abs/1507.05200}{[arXiv:1507.05200 [hep-ph]]}.
  %%CITATION = doi:10.1016/j.physletb.2015.12.071;%%
  %92 citations counted in INSPIRE as of 07 Oct 2017


%\cite{Burns:2015dwa}
\bibitem{Burns:2015dwa}
  T.~J.~Burns,
  Phenomenology of P$_{c}$(4380)$^{+}$, P$_{c}$(4450)$^{+}$ and related states,
  \href{https://link.springer.com/article/10.1140\%2Fepja\%2Fi2015-15152-6}{Eur.\ Phys.\ J.\ A {\bf 51}, 152 (2015)},
%  doi:10.1140/epja/i2015-15152-6
  \href{https://arxiv.org/abs/1509.02460}{[arXiv:1509.02460 [hep-ph]]}.
  %%CITATION = doi:10.1140/epja/i2015-15152-6;%%
  %58 citations counted in INSPIRE as of 07 Oct 2017

%\cite{Shimizu:2016rrd}
\bibitem{Shimizu:2016rrd}
  Y.~Shimizu, D.~Suenaga and M.~Harada,
  Coupled channel analysis of molecule picture of $P_{c}(4380)$,
  \href{https://journals.aps.org/prd/abstract/10.1103/PhysRevD.93.114003}{Phys.\ Rev.\ D {\bf 93}, 114003 (2016)},
  %doi:10.1103/PhysRevD.93.114003
  \href{https://arxiv.org/abs/1603.02376}{[arXiv:1603.02376 [hep-ph]]}.
  %%CITATION = doi:10.1103/PhysRevD.93.114003;%%
  %7 citations counted in INSPIRE as of 13 Sep 2016

%\cite{Chen:2016heh}
\bibitem{Chen:2016heh}
  R.~Chen, X.~Liu and S.~L.~Zhu,
  Hidden-charm molecular pentaquarks and their charmed-strange partners,
  \href{https://doi.org/10.1016/j.nuclphysa.2016.04.012}{Nucl.\ Phys.\ A {\bf 954}, 406 (2016)},
%  doi:10.1016/j.nuclphysa.2016.04.012
  \href{https://arxiv.org/abs/1601.03233}{[arXiv:1601.03233 [hep-ph]]}.
  %%CITATION = doi:10.1016/j.nuclphysa.2016.04.012;%%
  %10 citations counted in INSPIRE as of 13 Sep 2016


%\cite{Huang:2015uda}
\bibitem{Huang:2015uda}
  H.~Huang, C.~Deng, J.~Ping and F.~Wang,
  Possible pentaquarks with heavy quarks,
  \href{https://link.springer.com/article/10.1140\%2Fepjc\%2Fs10052-016-4476-z}{Eur.\ Phys.\ J.\ C {\bf 76}, 624 (2016)},
%  doi:10.1140/epjc/s10052-016-4476-z
  \href{https://arxiv.org/abs/1510.04648}{[arXiv:1510.04648 [hep-ph]]}.
  %%CITATION = doi:10.1140/epjc/s10052-016-4476-z;%%
  %31 citations counted in INSPIRE as of 07 Oct 2017

%\cite{Chen:2016otp}
\bibitem{Chen:2016otp}
  H.~X.~Chen, E.~L.~Cui, W.~Chen, X.~Liu, T.~G.~Steele and S.~L.~Zhu,
  QCD sum rule study of hidden-charm pentaquarks,
  \href{https://link.springer.com/article/10.1140\%2Fepjc\%2Fs10052-016-4438-5}{Eur.\ Phys.\ J.\ C {\bf 76}, 572 (2016)}
 % doi:10.1140/epjc/s10052-016-4438-5
  \href{https://arxiv.org/abs/1602.02433}{[arXiv:1602.02433 [hep-ph]]}.
  %%CITATION = doi:10.1140/epjc/s10052-016-4438-5;%%
  %15 citations counted in INSPIRE as of 07 Oct 2017

%\cite{Yang:2015bmv}
\bibitem{Yang:2015bmv}
  G.~Yang and J.~Ping,
  The structure of pentaquarks $P_c^+$ in the chiral quark model,
  \href{https://journals.aps.org/prd/abstract/10.1103/PhysRevD.95.014010}{Phys.\ Rev.\ D {\bf 95}, 014010 (2017)},
 % doi:10.1103/PhysRevD.95.014010
  \href{https://arxiv.org/abs/1511.09053}{[arXiv:1511.09053 [hep-ph]]}.
  %%CITATION = doi:10.1103/PhysRevD.95.014010;%%
  %18 citations counted in INSPIRE as of 07 Oct 2017

%\cite{He:2016pfa}
\bibitem{He:2016pfa}
  J.~He,
  Understanding spin parity of $P_c(4450)$ and $Y(4274)$ in a hadronic molecular state picture,
  \href{https://journals.aps.org/prd/abstract/10.1103/PhysRevD.95.074004}{Phys.\ Rev.\ D {\bf 95}, 074004 (2017)},
%  doi:10.1103/PhysRevD.95.074004
  \href{https://arxiv.org/abs/1607.03223}{[arXiv:1607.03223 [hep-ph]]}.
  %%CITATION = doi:10.1103/PhysRevD.95.074004;%%
  %9 citations counted in INSPIRE as of 07 Oct 2017

%\cite{Yamaguchi:2016ote}
\bibitem{Yamaguchi:2016ote}
  Y.~Yamaguchi and E.~Santopinto,
  Hidden-charm pentaquarks as a meson-baryon molecule with coupled channels for $\bar{D}^{(\ast)}\Lambda_{\rm c}$ and $\bar{D}^{(\ast)}\Sigma^{(\ast)}_{\rm c}$,
  \href{https://journals.aps.org/prd/abstract/10.1103/PhysRevD.96.014018}{Phys.\ Rev.\ D {\bf 96}, 014018 (2017)}
%  doi:10.1103/PhysRevD.96.014018
  \href{https://arxiv.org/abs/1606.08330}{[arXiv:1606.08330 [hep-ph]]}.
  %%CITATION = doi:10.1103/PhysRevD.96.014018;%%
  %16 citations counted in INSPIRE as of 07 Oct 2017

%\cite{Yamaguchi:2017zmn}
\bibitem{Yamaguchi:2017zmn}
  Y.~Yamaguchi, A.~Giachino, A.~Hosaka, E.~Santopinto, S.~Takeuchi and M.~Takizawa,
  Hidden-charm and bottom meson-baryon molecules coupled with five-quark states,
  \href{https://arxiv.org/abs/1709.00819}{arXiv:1709.00819 [hep-ph]}.
  %%CITATION = ARXIV:1709.00819;%%

%\cite{Xiao:2016mho}
\bibitem{Xiao:2016mho}
  C.~J.~Xiao and D.~Y.~Chen,
  Possible $B^{(\ast)} \bar{K}$ hadronic molecule state,
  \href{https://link.springer.com/article/10.1140\%2Fepja\%2Fi2017-12310-x}{Eur.\ Phys.\ J.\ A {\bf 53},  127 (2017)},
%  doi:10.1140/epja/i2017-12310-x
  \href{https://arxiv.org/abs/1603.00228}{[arXiv:1603.00228 [hep-ph]]}.
  %%CITATION = doi:10.1140/epja/i2017-12310-x;%%
  %24 citations counted in INSPIRE as of 12 Sep 2017

%\cite{Agaev:2016urs}
\bibitem{Agaev:2016urs}
  S.~S.~Agaev, K.~Azizi and H.~Sundu,
  Exploring $X(5568)$ as a meson molecule,
  \href{https://link.springer.com/article/10.1140\%2Fepjp\%2Fi2016-16351-8}{Eur.\ Phys.\ J.\ Plus {\bf 131}, 351 (2016)},
 % doi:10.1140/epjp/i2016-16351-8
  \href{https://arxiv.org/abs/1603.02708}{[arXiv:1603.02708 [hep-ph]]}.
  %%CITATION = doi:10.1140/epjp/i2016-16351-8;%%
  %32 citations counted in INSPIRE as of 12 Sep 2017

%\cite{Burns:2016gvy}
\bibitem{Burns:2016gvy}
  T.~J.~Burns and E.~S.~Swanson,
  Interpreting the $X(5568)$,
  \href{http://www.sciencedirect.com/science/article/pii/S0370269316303902?via\%3Dihub}{Phys.\ Lett.\ B {\bf 760}, 627 (2016)},
 % doi:10.1016/j.physletb.2016.07.049
  \href{https://arxiv.org/abs/1603.04366}{[arXiv:1603.04366 [hep-ph]]}.
  %%CITATION = doi:10.1016/j.physletb.2016.07.049;%%
  %38 citations counted in INSPIRE as of 12 Sep 2017

%\cite{Albaladejo:2016eps}
\bibitem{Albaladejo:2016eps}
  M.~Albaladejo, J.~Nieves, E.~Oset, Z.~F.~Sun and X.~Liu,
  Can $X(5568)$ be described as a $B_s\pi$, $B\bar{K}$ resonant state?,
  \href{http://www.sciencedirect.com/science/article/pii/S037026931630106X?via\%3Dihub}{Phys.\ Lett.\ B {\bf 757}, 515 (2016)},
 % doi:10.1016/j.physletb.2016.04.033
  \href{https://arxiv.org/abs/1603.09230}{[arXiv:1603.09230 [hep-ph]]}.
  %%CITATION = doi:10.1016/j.physletb.2016.04.033;%%
  %29 citations counted in INSPIRE as of 12 Sep 2017

%\cite{Chen:2016ypj}
\bibitem{Chen:2016ypj}
  R.~Chen and X.~Liu,
  Is the newly reported $X(5568)$ a $B\bar{K}$ molecular state?,
  \href{https://journals.aps.org/prd/abstract/10.1103/PhysRevD.94.034006}{Phys.\ Rev.\ D {\bf 94}, 034006 (2016)},
 % doi:10.1103/PhysRevD.94.034006
  \href{https://arxiv.org/abs/1607.05566}{[arXiv:1607.05566 [hep-ph]]}.
  %%CITATION = doi:10.1103/PhysRevD.94.034006;%%
  %10 citations counted in INSPIRE as of 12 Sep 2017

%\cite{Lu:2016kxm}
\bibitem{Lu:2016kxm}
  J.~X.~Lu, X.~L.~Ren and L.~S.~Geng,
  $B_s\pi - B\bar{K}$ interactions in finite volume and $X(5568)$,
  \href{https://link.springer.com/article/10.1140\%2Fepjc\%2Fs10052-017-4660-9}{Eur.\ Phys.\ J.\ C {\bf 77}, 94 (2017)},
 % doi:10.1140/epjc/s10052-017-4660-9
  \href{https://arxiv.org/abs/1607.06327}{[arXiv:1607.06327 [hep-ph]]}.
  %%CITATION = doi:10.1140/epjc/s10052-017-4660-9;%%
  %11 citations counted in INSPIRE as of 12 Sep 2017

%\cite{Sun:2016tmz}
\bibitem{Sun:2016tmz}
  B.~X.~Sun, F.~Y.~Dong and J.~L.~Pang,
  Study of $X(5568)$ in a unitary coupled-channel approximation of $B \bar{K}$ and $B_s \pi$,
  \href{http://iopscience.iop.org/article/10.1088/1674-1137/41/7/074104/meta;jsessionid=7014DA6EFBCF4DDA99D26039F0BF5562.c1.iopscience.cld.iop.org}{Chin.\ Phys.\ C {\bf 41}, 074104 (2017)},
 % doi:10.1088/1674-1137/41/7/074104
  \href{https://arxiv.org/abs/1609.04068}{[arXiv:1609.04068 [nucl-th]]}.
  %%CITATION = doi:10.1088/1674-1137/41/7/074104;%%
  %5 citations counted in INSPIRE as of 12 Sep 2017

%\cite{Barnes:2003dj}
\bibitem{Barnes:2003dj}
  T.~Barnes, F.~E.~Close and H.~J.~Lipkin,
  Implications of a $DK$ molecule at 2.32-GeV,
  \href{https://journals.aps.org/prd/abstract/10.1103/PhysRevD.68.054006}{Phys.\ Rev.\ D {\bf 68}, 054006 (2003)},
 % doi:10.1103/PhysRevD.68.054006
  \href{https://arxiv.org/abs/hep-ph/0305025}{[arXiv:hep-ph/0305025]}.
  %%CITATION = doi:10.1103/PhysRevD.68.054006;%%
  %328 citations counted in INSPIRE as of 12 Sep 2017

%\cite{Zhang:2009pn}
\bibitem{Zhang:2009pn}
  D.~Zhang, Q.~Y.~Zhao and Q.~Y.~Zhang,
  A Study of S-wave $DK$ interactions in the chiral SU(3) quark model,
  \href{http://iopscience.iop.org/article/10.1088/0256-307X/26/9/091201/meta}{Chin.\ Phys.\ Lett.\  {\bf 26}, 091201 (2009)},
 % doi:10.1088/0256-307X/26/9/091201
  \href{https://arxiv.org/abs/0905.1804}{[arXiv:0905.1804 [nucl-th]]}.
  %%CITATION = doi:10.1088/0256-307X/26/9/091201;%%
  %5 citations counted in INSPIRE as of 12 Sep 2017

%\cite{Guo:2006fu}
\bibitem{Guo:2006fu}
  F.~K.~Guo, P.~N.~Shen, H.~C.~Chiang, R.~G.~Ping and B.~S.~Zou,
  Dynamically generated $0^+$ heavy mesons in a heavy chiral unitary approach,
  \href{http://www.sciencedirect.com/science/article/pii/S0370269306011221?via\%3Dihub}{Phys.\ Lett.\ B {\bf 641}, 278 (2006)},
  %doi:10.1016/j.physletb.2006.08.064
  \href{https://arxiv.org/abs/hep-ph/0603072}{[hep-ph/0603072]}.
  %%CITATION = doi:10.1016/j.physletb.2006.08.064;%%
  %147 citations counted in INSPIRE as of 21 Nov 2017


%\cite{Xie:2010zza}
\bibitem{Xie:2010zza}
  Z.~X.~Xie, G.~Q.~Feng and X.~H.~Guo,
  Analyzing $D_{s0}^*(2317)^+$ in the $DK$ molecule picture in the Beth-Salpeter approach,
  \href{https://journals.aps.org/prd/abstract/10.1103/PhysRevD.81.036014}{Phys.\ Rev.\ D {\bf 81}, 036014 (2010)}.
 % doi:10.1103/PhysRevD.81.036014
  %%CITATION = doi:10.1103/PhysRevD.81.036014;%%
  %9 citations counted in INSPIRE as of 21 Nov 2017

%\cite{Navarra:2015iea}
\bibitem{Navarra:2015iea}
  F.~S.~Navarra, M.~Nielsen, E.~Oset and T.~Sekihara,
  Testing the molecular nature of $D_{s0}^*(2317)$ and $D_0^*(2400)$ in semileptonic $B_s$ and $B$ decays,
  \href{https://journals.aps.org/prd/abstract/10.1103/PhysRevD.92.014031}{Phys.\ Rev.\ D {\bf 92}, 014031 (2015},
 % doi:10.1103/PhysRevD.92.014031
  \href{https://arxiv.org/abs/1501.03422}{[arXiv:1501.03422 [hep-ph]]}.
  %%CITATION = doi:10.1103/PhysRevD.92.014031;%%
  %12 citations counted in INSPIRE as of 21 Nov 2017

%\cite{Guo:2011dd}
\bibitem{Guo:2011dd}
  F.~K.~Guo and U.~G.~Meissner,
  More kaonic bound states and a comprehensive interpretation of the $D_{sJ}$ states,
  \href{https://journals.aps.org/prd/abstract/10.1103/PhysRevD.84.014013}{Phys.\ Rev.\ D {\bf 84}, 014013 (2011)},
  %doi:10.1103/PhysRevD.84.014013
  \href{https://arxiv.org/abs/1102.3536}{[arXiv:1102.3536 [hep-ph]]}.
  %%CITATION = doi:10.1103/PhysRevD.84.014013;%%
  %25 citations counted in INSPIRE as of 12 Sep 2017

%\cite{Guo:2006rp}
\bibitem{Guo:2006rp}
  F.~K.~Guo, P.~N.~Shen and H.~C.~Chiang,
  Dynamically generated $1^+$ heavy mesons,
  \href{http://www.sciencedirect.com/science/article/pii/S0370269307001724?via\%3Dihub}{Phys.\ Lett.\ B {\bf 647}, 133 (2007)},
 % doi:10.1016/j.physletb.2007.01.050
  \href{https://arxiv.org/abs/hep-ph/0610008}{[hep-ph/0610008]}.
  %%CITATION = doi:10.1016/j.physletb.2007.01.050;%%
  %80 citations counted in INSPIRE as of 21 Nov 2017


%\cite{Faessler:2007us}
\bibitem{Faessler:2007us}
  A.~Faessler, T.~Gutsche, V.~E.~Lyubovitskij and Y.~L.~Ma,
  $D^* K$ molecular structure of the $D_{s1}(2460)$ meson,
  \href{https://journals.aps.org/prd/abstract/10.1103/PhysRevD.76.114008}{Phys.\ Rev.\ D {\bf 76}, 114008 (2007)},
 % doi:10.1103/PhysRevD.76.114008
  \href{https://arxiv.org/abs/0709.3946}{[arXiv:0709.3946 [hep-ph]]}.
  %%CITATION = doi:10.1103/PhysRevD.76.114008;%%
  %73 citations counted in INSPIRE as of 12 Sep 2017

%\cite{Feng:2012zze}
\bibitem{Feng:2012zze}
  G.~Q.~Feng, X.~H.~Guo and Z.~H.~Zhang,
  Studying the $D^* K$ molecular structure of $D_s(2460)$ in the Bethe-Salpeter approach,
  \href{https://link.springer.com/article/10.1140\%2Fepjc\%2Fs10052-012-2033-y}{Eur.\ Phys.\ J.\ C {\bf 72}, 2033 (2012)}.
%  doi:10.1140/epjc/s10052-012-2033-y
  %%CITATION = doi:10.1140/epjc/s10052-012-2033-y;%%
  %3 citations counted in INSPIRE as of 21 Nov 2017


%\cite{Weinstein:1982gc}
\bibitem{Weinstein:1982gc}
  J.~D.~Weinstein and N.~Isgur,
  Do Multi-Quark Hadrons Exist?,
  \href{https://journals.aps.org/prl/abstract/10.1103/PhysRevLett.48.659}{Phys.\ Rev.\ Lett.\  {\bf 48}, 659 (1982)}.  %%CITATION = PRLTA,48,659

%\cite{Weinstein:1983gd}
\bibitem{Weinstein:1983gd}
  J.~D.~Weinstein and N.~Isgur,
  The q q anti-q anti-q System in a Potential Model,
  \href{https://journals.aps.org/prd/abstract/10.1103/PhysRevD.27.588}{Phys.\ Rev.\ D {\bf 27}, 588 (1983)}.  %%CITATION = PHRVA,D27,588;%%

%\cite{Weinstein:1990gu}
\bibitem{Weinstein:1990gu}
  J.~D.~Weinstein and N.~Isgur,
  K anti-K Molecules,
  \href{https://journals.aps.org/prd/abstract/10.1103/PhysRevD.41.2236}{Phys.\ Rev.\ D {\bf 41}, 2236 (1990)}.  %%CITATION = PHRVA,D41,2236;%%


%\cite{Roca:2005nm}
\bibitem{Roca:2005nm}
  L.~Roca, E.~Oset and J.~Singh,
  Low lying axial-vector mesons as dynamically generated resonances,
  \href{https://journals.aps.org/prd/abstract/10.1103/PhysRevD.72.014002}{Phys.\ Rev.\ D {\bf 72}, 014002 (2005)},
  %doi:10.1103/PhysRevD.72.014002
  \href{https://arxiv.org/abs/hep-ph/0503273}{[hep-ph/0503273]}.
  %%CITATION = doi:10.1103/PhysRevD.72.014002;%%
  %150 citations counted in INSPIRE as of 21 Nov 2017

%\cite{Aceti:2015zva}
\bibitem{Aceti:2015zva}
  F.~Aceti, J.~M.~Dias and E.~Oset,
  $f_{1}(1285)$ decays into $a_{0}(980)\pi^{0}$, $f_{0}(980)\pi^{0}$ and isospin breaking,
  \href{https://link.springer.com/article/10.1140\%2Fepja\%2Fi2015-15048-5}{Eur.\ Phys.\ J.\ A {\bf 51}, 48 (2015)},
  %doi:10.1140/epja/i2015-15048-5
  \href{https://arxiv.org/abs/1501.06505}{[arXiv:1501.06505 [hep-ph]]}.
  %%CITATION = doi:10.1140/epja/i2015-15048-5;%%
  %28 citations counted in INSPIRE as of 21 Nov 2017

%\cite{Geng:2015yta}
\bibitem{Geng:2015yta}
  L.~S.~Geng, X.~L.~Ren, Y.~Zhou, H.~X.~Chen and E.~Oset,
  $S$-wave $KK^*$ interactions in a finite volume and the $f_1(1285)$,
  \href{http://dx.doi.org/10.1103/PhysRevD.92.014029}{Phys.\ Rev.\ D {\bf 92}, 014029 (2015)},
 % doi:10.1103/PhysRevD.92.014029
  \href{https://arxiv.org/abs/1503.06633}{[arXiv:1503.06633 [hep-ph]]}.
  %%CITATION = doi:10.1103/PhysRevD.92.014029;%%
  %11 citations counted in INSPIRE as of 21 Nov 2017


%\cite{Lu:2016nlp}
\bibitem{Lu:2016nlp}
  P.~L.~L¨¹ and J.~He,
  Hadronic molecular states from the $K\bar{K}^{\ast}$ interaction,
  \href{https://link.springer.com/article/10.1140\%2Fepja\%2Fi2016-16359-7}{Eur.\ Phys.\ J.\ A {\bf 52},  359 (2016)},
 % doi:10.1140/epja/i2016-16359-7
  \href{https://arxiv.org/abs/1603.04168}{[arXiv:1603.04168 [hep-ph]]}.
  %%CITATION = doi:10.1140/epja/i2016-16359-7;%%
  %5 citations counted in INSPIRE as of 12 Sep 2017

%\cite{Dalitz:1959dn}
\bibitem{Dalitz:1959dn}
  R.~H.~Dalitz and S.~F.~Tuan,
  A possible resonant state in pion-hyperon scattering,
  \href{https://journals.aps.org/prl/abstract/10.1103/PhysRevLett.2.425}{Phys.\ Rev.\ Lett.\  {\bf 2}, 425 (1959)}.
 % doi:10.1103/PhysRevLett.2.425
  %%CITATION = doi:10.1103/PhysRevLett.2.425;%%
  %140 citations counted in INSPIRE as of 12 Sep 2017

%\cite{Dalitz:1960du}
\bibitem{Dalitz:1960du}
  R.~H.~Dalitz and S.~F.~Tuan,
  The phenomenological description of $K-$nucleon reaction processes,
  \href{http://www.sciencedirect.com/science/article/pii/0003491660900014?via\%3Dihub}{Annals Phys.\  {\bf 10}, 307 (1960)}.
 % doi:10.1016/0003-4916(60)90001-4
  %%CITATION = doi:10.1016/0003-4916(60)90001-4;%%
  %330 citations counted in INSPIRE as of 12 Sep 2017

%\cite{Ezoe:2016mkp}
\bibitem{Ezoe:2016mkp}
  T.~Ezoe and A.~Hosaka,
  Kaon-Nucleon systems and their interactions in the Skyrme model,
  \href{https://journals.aps.org/prd/abstract/10.1103/PhysRevD.94.034022}{Phys.\ Rev.\ D {\bf 94}, 034022 (2016)},
%  doi:10.1103/PhysRevD.94.034022
  \href{https://arxiv.org/abs/1605.01203}{[arXiv:1605.01203 [nucl-th]]}.
  %%CITATION = doi:10.1103/PhysRevD.94.034022;%%
  %2 citations counted in INSPIRE as of 12 Sep 2017

%\cite{Aaij:2017nav}
\bibitem{Aaij:2017nav}
  R.~Aaij {\it et al.} [LHCb Collaboration],
  Observation of five new narrow $\Omega_c^0$ states decaying to $\Xi_c^+ K^-$,
  \href{https://journals.aps.org/prl/abstract/10.1103/PhysRevLett.118.182001}{Phys.\ Rev.\ Lett.\  {\bf 118}, 182001 (2017)},
 % doi:10.1103/PhysRevLett.118.182001
  \href{https://arxiv.org/abs/1703.04639}{[arXiv:1703.04639 [hep-ex]]}.
  %%CITATION = doi:10.1103/PhysRevLett.118.182001;%%
  %33 citations counted in INSPIRE as of 13 Sep 2017

%\cite{Kim:2017jpx}
\bibitem{Kim:2017jpx}
  H.~C.~Kim, M.~V.~Polyakov and M.~Praszalowicz,
  Possibility of the existence of charmed exotica,
  \href{https://journals.aps.org/prd/abstract/10.1103/PhysRevD.96.014009}{Phys.\ Rev.\ D {\bf 96}, 014009 (2017)
  Addendum: [Phys.\ Rev.\ D {\bf 96}, 039902 (2017)]},
%  doi:10.1103/PhysRevD.96.039902, 10.1103/PhysRevD.96.014009
  \href{https://arxiv.org/abs/1704.04082}{[arXiv:1704.04082 [hep-ph]]}.
  %%CITATION = doi:10.1103/PhysRevD.96.039902, 10.1103/PhysRevD.96.014009;%%
  %10 citations counted in INSPIRE as of 13 Sep 2017

%\cite{Yang:2017rpg}
\bibitem{Yang:2017rpg}
  G.~Yang and J.~Ping,
  The structure of pentaquarks $\Omega_c^0$ in the chiral quark model,
  \href{https://arxiv.org/abs/1703.08845}{arXiv:1703.08845 [hep-ph]}.
  %%CITATION = ARXIV:1703.08845;%%
  %13 citations counted in INSPIRE as of 13 Sep 2017

%\cite{Huang:2017dwn}
\bibitem{Huang:2017dwn}
  H.~Huang, J.~Ping and F.~Wang,
  Investigating the excited $\Omega^{0}_{c}$ states through $\Xi_{c}K$ and $\Xi^{'}_{c}K$ decay channels,
  \href{https://arxiv.org/abs/1704.01421}{arXiv:1704.01421 [hep-ph]}.
  %%CITATION = ARXIV:1704.01421;%%
  %11 citations counted in INSPIRE as of 13 Sep 2017

%\cite{Montana:2017kjw}
\bibitem{Montana:2017kjw}
  G.~Montana, A.~Feijoo and A.~Ramos,
  A meson-baryon molecular interpretation for some $\Omega_c$ excited baryons,
  \href{https://arxiv.org/abs/1709.08737}{arXiv:1709.08737 [hep-ph]}.
  %%CITATION = ARXIV:1709.08737;%%

%\cite{Debastiani:2017ewu}
\bibitem{Debastiani:2017ewu}
  V.~R.~Debastiani, J.~M.~Dias, W.~H.~Liang and E.~Oset,
  Molecular $\Omega_c$ states within the local hidden gauge approach,
  \href{https://arxiv.org/abs/1710.04231}{arXiv:1710.04231 [hep-ph]}.
  %%CITATION = ARXIV:1710.04231;%%

%\cite{Liu:2011xc}
\bibitem{Liu:2011xc}
  Y.~R.~Liu and M.~Oka,
  $\Lambda_c N$ bound states revisited,
  \href{https://journals.aps.org/prd/abstract/10.1103/PhysRevD.85.014015}{Phys.\ Rev.\ D {\bf 85}, 014015 (2012)},
  \href{https://arxiv.org/abs/1103.4624}{[arXiv:1103.4624 [hep-ph]]}.
  %%CITATION = ARXIV:1103.4624;%%
  %15 citations counted in INSPIRE as of 17 juin 2015

%\cite{Lin:1999ad}
\bibitem{Lin:1999ad}
  Z.~-w.~Lin and C.~M.~Ko,
  A Model for J / psi absorption in hadronic matter,
  \href{https://journals.aps.org/prc/abstract/10.1103/PhysRevC.62.034903}{Phys.\ Rev.\ C {\bf 62}, 034903 (2000)}
  \href{https://arxiv.org/abs/nucl-th/9912046}{[arXiv:nucl-th/9912046]}.
  %%CITATION = NUCL-TH/9912046;%%

%\cite{Nagahiro:2008mn}
\bibitem{Nagahiro:2008mn}
  H.~Nagahiro, L.~Roca and E.~Oset,
  Meson loops in the $f_0(980)$ and $a_0(980)$ radiative decays into $\rho$, $\omega$,
  \href{https://link.springer.com/article/10.1140\%2Fepja\%2Fi2008-10567-8}{Eur.\ Phys.\ J.\ A {\bf 36}, 73 (2008)},
%  doi:10.1140/epja/i2008-10567-8
  \href{https://arxiv.org/abs/0802.0455}{[arXiv:0802.0455 [hep-ph]]}.
  %%CITATION = doi:10.1140/epja/i2008-10567-8;%%
  %30 citations counted in INSPIRE as of 13 Sep 2017

%\cite{Chen:2017vai}
\bibitem{Chen:2017vai}
  R.~Chen, X.~Liu and A.~Hosaka,
  Heavy molecules and one-$\sigma/\omega$-exchange model,
  \href{https://arxiv.org/abs/1707.08306}{arXiv:1707.08306 [hep-ph]}.
  %%CITATION = ARXIV:1707.08306;%%
  %1 citations counted in INSPIRE as of 18 Sep 2017

%\cite{Tornqvist:1993ng}
\bibitem{Tornqvist:1993ng}
  N.~A.~Tornqvist,
From the Deuteron to Deusons, an Analysis of Deuteron-like Meson Meson Bound States,
  \href{https://link.springer.com/article/10.1007\%2FBF01413192}{Z.\ Phys.\ C {\bf 61}, 525 (1994)},
  \href{https://arxiv.org/abs/hep-ph/9310247}{[arXiv:hep-ph/9310247]}.
  %%CITATION = HEP-PH/9310247;%%
  %250 citations counted in INSPIRE as of 22 Oct 2015

%\cite{Tornqvist:1993vu}
\bibitem{Tornqvist:1993vu}
  N.~A.~Tornqvist,
 On Deusons or Deuteron-like Meson Meson Bound States,
  \href{https://link.springer.com/article/10.1007\%2FBF02734018}{Nuovo Cim.\ A {\bf 107}, 2471 (1994)},
  \href{https://arxiv.org/abs/hep-ph/9310225}{[arXiv:hep-ph/9310225]}.
  %%CITATION = HEP-PH/9310225;%%
  %27 citations counted in INSPIRE as of 22 Oct 2015


%\cite{Klempt:2002ap}
\bibitem{Klempt:2002ap}
  E.~Klempt, F.~Bradamante, A.~Martin, and J.~M.~Richard,
  Antinucleon nucleon interaction at low energy: Scattering and protonium,
  \href{http://www.sciencedirect.com/science/article/pii/S0370157302001448?via\%3Dihub}{Phys.\ Rep.\  {\bf 368}, 119 (2002)}.
 % doi:10.1016/S0370-1573(02)00144-8
  %%CITATION = doi:10.1016/S0370-1573(02)00144-8;%%
  %107 citations counted in INSPIRE as of 20 Feb 2017





\end{thebibliography}
\end{document}